\documentclass[final,3p]{elsarticle}

\DeclareMathAlphabet{\pazocal}{OMS}{zplm}{m}{n}
\DeclareMathAlphabet{\bpazocal}{OMS}{cmsy}{b}{n}

\RequirePackage[ruled,vlined]{algorithm2e}

\SetCommentSty{xCommentSty}
\RequirePackage{caption,subcaption}

\RequirePackage{lineno}
\RequirePackage{xcolor}
\RequirePackage{amsmath,amsfonts,amsthm,amssymb}

\RequirePackage[breaklinks]{hyperref}
\RequirePackage{stmaryrd}

\usepackage[capitalise]{cleveref}
\RequirePackage{cleveref}

\RequirePackage{tikz}
\RequirePackage{pgfplots}
\RequirePackage{multicol}\RequirePackage{multirow}

\usepackage{comment} \usepackage{todonotes}

\usetikzlibrary{external}
\usetikzlibrary{arrows.meta}
\usetikzlibrary{shapes.geometric}
\usetikzlibrary{matrix,automata,fit,calc,positioning}

\pgfplotsset{compat=newest}
\usepgfplotslibrary{fillbetween}
\usepgfplotslibrary{groupplots}
\usepgfplotslibrary{external}
\usetikzlibrary{patterns}

\tikzset{tipA/.tip={Triangle[angle=45:4pt]Bar[width=5.0mm]},
  tipB/.tip={Triangle[angle=45:4pt]} ,
  tipC/.tip={Bar[width=5.0mm]},
  tipD/.tip={Bar[width=2.5mm]},
}
\tikzset{external/only named=true}

\definecolor{blackmy}{RGB}{38, 70, 83}
\definecolor{bluemy}{RGB}{39, 125, 161}
\definecolor{greenmy}{RGB}{42, 167, 143}
\definecolor{yellowmy}{RGB}{233, 196, 106}
\definecolor{brownmy}{RGB}{244, 162, 97}
\definecolor{redmy}{RGB}{249, 65, 68}
\definecolor{cyanmy}{RGB}{5, 255, 255}
\definecolor{purplemy}{RGB}{129, 80, 192}
\tikzset{external/up to date check=md5}\renewcommand{\vec}[1]{\boldsymbol{#1}}

\def \sv{\vec{s}}

\def \uv{\vec{u}}

\def \xv{\vec{x}}

\def \R{\mathbb{R}}

\makeatletter
\newcommand{\nosemic}{\renewcommand{\@endalgocfline}{\relax}}\newcommand{\dosemic}{\renewcommand{\@endalgocfline}{\algocf@endline}}\let\oldnl\nl \newcommand{\nonl}{\renewcommand{\nl}{\let\nl\oldnl}}\makeatother

\journal{Computer Methods in Applied Mechanics and Engineering}
\usepackage{tikz}
\usepackage{pgfplots}
\usepackage{pgfplotstable}

\pgfplotsset{compat=1.17}

\begin{document}

\begin{frontmatter}

\title{The limitations of a standard phase-field model in reproducing jointing in sedimentary rock layers}

\author[ETH]{E. Pezzulli\corref{myself}}
\address[ETH]{Department of Earth Sciences, ETH Zurich, Switzerland}
\cortext[myself]{Department of Earth Sciences, ETH Zurich, Switzerland }
\ead{edoardo.pezzulli@erdw.ethz.ch}

\author[USI,UniDistance]{P. Zulian}
\author[USI]{ A. Kopani\v{c}\'{a}kov\'{a}}
\author[USI,UniDistance]{ R. Krause}

\address[USI]{Euler Institute, Universit\`{a} della Svizzera italiana, Lugano, Switzerland}
\address[UniDistance]{UniDistance Suisse, Brig, Switzerland}

\author[ETH]{T. Driesner}

  \begin{abstract}
    
Geological applications of phase-field methods for fracture are notably scarce. This work conducts a numerical examination of the applicability of standard phase-field models in reproducing jointing within sedimentary layers. We explore how the volumetric-deviatoric split alongside the AT1 and AT2 phase-field formulations have several advantages in simulating jointing, but also have intrinsic limitations that prevent a reliable quantitative analysis of rock fracture. The formulations qualitatively reproduce the process of joint saturation, along with the negative correlation between joint spacing and the height of the sedimentary layer. However, in quantitative comparison to alternative numerical methods and outcrop observations, the phase-field method overestimates joint spacings by a factor of 2 and induces unrealistic compressive fractures in the AT1 model, alongside premature shearing at layer interfaces for the AT2 model. The causes are identified to be intrinsic to the phase-field lengthscale and the unsuitable strength envelope arising from the Volumetric-Deviatoric split. Finally, our analysis elucidates on the phase-field lengthscale's distortion of the stress field around dilating fractures, causing the sedimentary layer to reach joint saturation prematurely, thereby stopping the nucleation of new fractures and leading to larger joint spacings than in natural examples. Decreasing the lengthscale results in gradual improvement but becomes prohibitively computationally expensive for very small lengthscales such that the limit of ``natural'' behavior was not reached in this study. Overall, our results motivate the development of constitutive phase-field models that are more suitable for geological applications and their benchmarking against geological observations.

   \end{abstract}

\end{frontmatter}

\begin{sloppypar}
  \section{Introduction}
\label{sec:introduction}
Fractured geothermal reservoirs have the potential to provide an important source of economically-viable geothermal energy. The productivity of a fractured geothermal reservoir is heavily dependent on the permeability provided by the reservoirs natural fractures, as the host rock is typically too impermeable to sustain economic rates of fluid production. Therefore, characterising a geothermal reservoir's fracture system, and it's associated permeability enhancement, constitutes an essential approach to assess the economic potential of a fractured geothermal system.    
\hfill\newline

Characterising a naturally fractured geothermal reservoir remains a very challenging task. Observations of fracture networks are notoriously difficult to obtain. Borehole tools, such as acoustic and optical televiewers, are capable of inferring the density and orientations of fractures that intersect the wellbore, while borehole ground penetrating radar is only recently offering glimpses several decameters beyond the borehole walls \cite{shakas_permeability_2020}. Outcrop studies inform on $2D$ fracture geometries, but often lack the 3-dimensional perspective and are overprinted by the more recent history of the reservoir \cite{bauer_predictability_2017}. Geophysical techniques such as active seismic, attempt to infer fracture densities and orientations from the change in seismic velocities and stress anisotropy caused by fractures \cite{yousef_when_2016, angus_reservoir_2016}. This relies on adequate rock physics models and the cost of seismic campaigns which must reconcile with the limited exploration budgets of a geothermal project. Consequently, very little is known about the fracture lengths, densities, and orientations of fractures within a reservoir. As a result, numerical modelling constitutes an important tool to constrain the reservoir's fracture network. 
\hfill\newline

To simulate the generation of a fracture network,  numerical methods are confronted with the formidable challenge of simulating fracture nucleation, propagation, and coalescence. Very few numerical methods have the capacity to handle all three complex phenomena in a single model. Extensive work over the last two decades has seen the wide application of the finite-discrete element method (FDEM) \cite{munjiza_combined_2004, lisjak_review_2014, lei_use_2017, mohammadnejad_overview_2021, pezzulli_finite_2022,pezzulli_enhanced_2022,pezzulli_energy-based_2022} within the geotechnical and petroleum industries \cite{elmo_applications_2013, mahabadi_y-geo_2012, mahabadi_development_2016}. Such methods can accurately capture a fracture's displacement discontinuity and its resultant stress perturbations, but typically require extensive remeshing techniques to track the evolving fracture surfaces. Furthermore, non-linear contact detection and interaction algorithms must be deployed to handle contact, friction and slip. Fracture propagation is not handled implicitly, but must be evaluated within an external non-linear iteration loop \cite{lecampion_numerical_2018, salimzadeh_three-dimensional_2018, paluszny_finite-element_2018}. As a result, such methods face significant challenges in resolving the competition of multiple propagating fracture tips, with different model parameterisations leading to distinctly different final fracture network configurations \cite{renshaw_numerical_1994, olson_joint_1993,thomas_growth_2020}. Overall, despite the advances and adoption of such methods in the industry, alternatives with lower algorithmic and computational complexity are highly desired in order to generate “geomechanically realistic” fracture networks.
\hfill\newline

The phase-field method appears to offer an elegant solution to handle fracture nucleation, propagation, and coalescence. The elegance stems from the reformulation of the fracture mechanical problem from a variational perspective of energy minimisation, allowing nucleation and propagation to be considered simultaneously within the same framework \cite{francfort_revisiting_1998,bourdin_variational_2008}. Another essential advantage stems from the regularisation of the fracture. This enables a fracture surface to be approximated by a crack surface \textit{density} which depends on a scalar damage (phase) field \cite{ambrosio_approximation_1990,kristensen_assessment_2021}. Consequently, fracture surfaces no longer need to be tracked, but can be inferred and evolved implicitly through a damage-evolution equation. The technique has been extended to include thermo-elastic \cite{miehe_phase_2016}, poro-elastic \cite{jammoul_phase-field-based_2022},  conchoidal~\cite{bilgen2018phase},  dynamic \cite{schluter_phase_2014,li_gradient-damage_2016} and ductile fracture \cite{ambati_phase-field_2015}, with excellent reviews given by \cite{wu_chapter_2020,de_lorenzis_numerical_2020}. The relative `ease' in generating fracture networks with phase-field models has compelled researchers to apply such methods in geological applications.
\hfill\newline

However, significant complexities arise when reproducing geological fracture networks. Fractures in geological media form i) under both tension and compression, ii) within quasi-brittle rocks that display significant mode-dependent fracture toughness, and iii) material layering that control the propagation, termination and deflection of natural fractures at material interfaces. Each of these processes poses significant challenges to the phase-field method. Fracture nucleation in compressive stress-fields is an active area of research, with differing schools of thought on which phase-field formulation can adequately reproduce experimentally-observed strength envelopes  \cite{kumar_revisiting_2020,de_lorenzis_nucleation_2022, navidtehrani_general_2022, vicentini_energy_2023}. Regarding mode-dependent failure, a common observation for rocks is that mode II fracture toughness can be 10 times greater than failure under mode I \cite{bahrami_theory_2020}. Yet, current phase-fields models are challenged with an intrinsic difficulty in distinguishing which part of the elastic energy is responsible for each failure mode. Several `energy splits' have been proposed to address such an issue based on deviatoric or shear energies \cite{zhang_modification_2017,liu_thermodynamically_2022}, introducing multiple damage (phase) fields \cite{bleyer_phase-field_2018,fei_double-phase-field_2021}, or leveraging the directional dependence of shear fractures \cite{bryant_mixed-mode_2018,sun_phase-field_2023}. However, most solutions break the variational consistency of the original formulation, and their ad-hoc nature require extensive validation against real applications. Finally, the regularisation lengthscale inherent to phase-field models indirectly imposes constraints on the rate at which material properties must vary if they are to be `seen' by the fracture \cite{vicentini_phase-field_2023}.  Consequently, it remains unclear how the regularisation lengthscale inherent to phase-field models is to be reconciled with sharp material boundaries and interfaces \cite{nguyen_role_2019,paggi_revisiting_2017, reinoso_crack_2019}. Overall, the verdict is not yet out on which phase-field models reliably reproduce fracture patterns commonly observed in the subsurface. 
\hfill\newline

\begin{figure}
    \centering
    \includegraphics[width=0.7\textwidth]{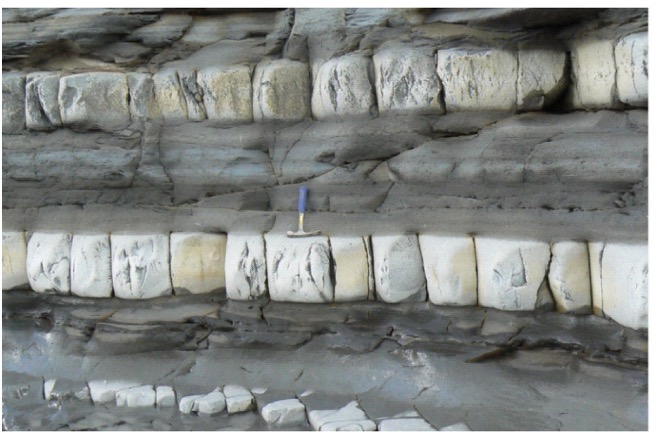}
    \caption{Field examples of jointing; layered and fractured (jointed) sedimentary rocks modified from Schöpfer et al. (2011) \cite{schopfer_reconciliation_2011} and Chemanda et al (2021) \cite{chemenda_origin_2021}. Alternating incompetent (mudrocks) and fractured competent (limestone) layers in Lilstock Bay, Somerset, UK. Fracture spacing has been observed to be proportional to the height of the competent layer.}
    \label{fig:field_jointing}
\end{figure}

There is a strong need to evaluate the behaviour of standard phase-field models for fracture within practical geological contexts. Yet, geological applications of phase-field methods are currently extremely rare. A particularly relevant example regards the simulation of jointing within sedimentary layers, for which  a field example is shown in \cref{fig:field_jointing}. Jointing has been extensively studied using outcrop mapping \cite{pollard_progress_1988,helgeson_characteristics_1991,gross_factors_1995,cooke_fracture_2001,underwood_stratigraphic_2003}, analytical \cite{hobbs_formation_1967,yin_fracture_2010} and numerical models such as the finite-discrete element method \cite{bai_fracture_2000,bai_closely_2000}, discrete element method \cite{schopfer_reconciliation_2011}, damage mechanics \cite{tang_fracture_2008}, and elasto-plastic models \cite{chemenda_origin_2021}. As such, jointed sedimentary sequences provide a valuable validation-case with which to test phase-field models. Several phase-field studies have focused on the termination, deflection and propagation of joints at material interfaces \cite{nguyen_role_2019, hansen-dorr_phase-field_2019,hansen-dorr_phase-field_2020,liu_variational_2021, reinoso_crack_2019}. Insofar, only Chuckwudozie \textit{et.al} \cite{chukwudozie_new_2013} have studied the jointing process within extending sedimentary layers using an isotropic AT2 phase-field formulation. Consequently, a deeper analysis on the performance of phase-field models with regards to reproducing commonly-observed characteristics of jointing is warranted. 
\hfill\newline

This work aims to evaluate the geological-readiness of standard phase-field models in reproducing natural fracture patterns observed in sedimentary rocks. In particular, we assess the behaviour of the volumetric-deviatoric split alongside the AT1 and the AT2 model. The numerical analysis begins with an overview of the evolution of damage, stress and energy in \cref{sec:pf_perspective}, where phase-field predictions are consistent with the existing understanding of jointing, thereby suggesting the basic suitability of the method for the geological problem. The mechanics of the phase-field method are explained in detail, with simulation results offering insights on jointing processes such as joint saturation and joint dilation.  In \cref{sec:pf_challenges}, a critical evaluation is performed of some of the limitations of the current phase-field models when reproducing common characteristics of jointed sequences. The impact of the phase-field lengthscale and the consequence of simplified strength envelopes is elucidated on. We close the discussion in \cref{sec:conclusion} by illustrating the potential of the phase-field method through an example of complex joint nucleation and interaction in 3-dimensions. Overall, the study concludes by emphasizing ongoing efforts in developing more suitable constitutive models and calls for continued benchmarking against geological observations.

   \section{Methodology}
\label{sec:pf}

We consider the deformation of a linear elastic body capable of brittle failure under external loading. To this end, we define our elastic body ${\Omega \subset \R^d}$, ${d \in \{2,3\}}$, with boundary $\Gamma$ that is further decomposed into a Dirichlet boundary $\Gamma_D$ and a Neumann boundary $\Gamma_N = \Gamma \setminus \Gamma_D$, on which Dirichlet and Neumann boundary conditions are applied, respectively. We consider a quasi-static loading process wherby the Dirichlet and Neumann boundary conditions evolve with a pseudo time ${t= 1,\ldots, T}$. Due to the application of these boundary conditions, the body $\Omega$ undergoes deformation ${\uv:\Omega\to \R^d}$, which in turn causes crack propagation.  The distribution of cracks in the material is described by the scalar phase-field parameter $c\in[0,1]$, with the limits $c=0$ and $c=1$ denoting the undamaged (intact) and fully-damaged (broken) material phases, respectively.

\subsection{Governing equations}
 The variational approach to fracture \cite{francfort_revisiting_1998} describes the evolution of the mechanical system and its phase-field parameter $c$ through the minimisation of the energy functional 
\begin{linenomath*}
  \begin{equation}
    \pazocal{E}(\uv, c) := 
    \int_{\Omega} \underbrace{ g(c)\psi^{d}_e(\uv) + \psi^{r}_e(\uv)}_{\psi_e} \, d\Omega + 
    \int_\Omega \underbrace{G_c \gamma(c,\nabla c)}_{\psi_f} \,  d\Omega \, - \int_{\Gamma_N} \textbf{t} \cdot \uv d\Gamma \, .
    \label{eq:energy_regularized}
  \end{equation}
\end{linenomath*}
 The first integral on the right hand side describes the total elastic energy embodied in the rock. It is composed of an integral of the elastic energy density $\psi_e$, which is the sum of a residual energy density $\psi^r_e$ that is unaffected by damage, and a degradable energy density $\psi^d_e$ that is degraded by damage. The degradation can be interpreted as a softening of the material \cite{marigo_overview_2016}, and is performed through the action of a degradation function $g(c)$. The degradation function $g(c)$ reduces to zero for a fully-broken ($c=1$) material. The drop in elastic energy due to damage (and softening of the material) comes at the cost of an increase in fracture surface energy, expressed through the second integral on the right hand side of \cref{eq:energy_regularized}. The magnitude of the Griffith critical energy release rate $G_c$ \cite{griffith_vi_1921} determines the energetic `price' of fracturing, and is proportional to the fracture surface area $A=\int_\Omega \gamma\,\partial \Omega$, which is obtained upon integration of a crack surface \textit{density} $\gamma$ \cite{kristensen_assessment_2021}. The last integral in \cref{eq:energy_regularized} represents the external work done on the boundary of the body by external tractions \textbf{t}.
 \hfill\newline
 
 The rise in popularity of phase-field models is partly due to the discovery of a family of crack surface density functions 
\begin{equation}\label{eq:crack_surface_density}
    \gamma(c,\nabla C) = \frac{1}{4 c_\omega l}\big(\omega(c)+ l^2|\nabla c|^2\big)\,,
\end{equation}
that are proven to converge to the crack surface area $A$ for a sufficiently small lengthscale parameter $l$ \cite{ambrosio_approximation_1990}. Such crack surface density functions commonly depend on the crack dissipation function $\omega(c)$ with $\omega(0)=0$ and $\omega(1)=1$, and have a normalisation parameter $c_\omega=\int_0^1 \sqrt{\omega(c)}\,dc$. The phase-field problem can be solved by minimising the functional described in \cref{eq:energy_regularized}, while also ensuring the irreversibility of damage. However, the phase-field problem is commonly solved by formulating the strong form of the governing equations, termed Euler-Lagrange equations, that are derived by taking the \textit{functional} derivative of the energy functional \cref{eq:energy_regularized}. The Euler-Lagrange equations are 
\begin{subequations}\label{eq:strongform}
\begin{equation}
    \nabla \cdot \boldsymbol \sigma + b = 0 \, ,  
\end{equation}
\begin{equation}\label{eq:phase_field_evolution}
    w_1 \big( \omega'(c) \,- \,2\, l^2\, \nabla^2 c \big) + g'(c) \psi^d(\uv) = 0 \, , 
\end{equation}
\end{subequations}
where the stress is defined as
\begin{equation}\label{eq:stress_def}
    \sigma_{ij} = g(c)\frac{\partial \psi^d}{\partial\varepsilon_{ij}} + \frac{\partial \psi^r}{\partial\varepsilon_{ij}}\,,
\end{equation}
and $w_1=\frac{G_c}{4c_\omega l}$. The phase-field evolution described in \cref{eq:phase_field_evolution} can be loosely interpreted as a diffusive ($\nabla^2$ term) and a reaction term, which is being driven by the degraded elastic energy $g(c)\psi^d$ which acts as a source of damage $c$. 
\hfill\newline

To complete the formulation, the irreversible nature of fracture must be incorporated through the inequality constraint $\dot{c} \geq 0$. The symbol $\dot{c}$ represents the derivative of $c$ with respect to time. This irreversibility constraint must also be complemented with the fact that the phase-field parameter must stay bounded between $0$ and $1$; introducing additional inequality constraints which must be enforced during the solution process. 
\hfill\newline

\paragraph{Remark} A phase-field model is said to be variationally consistent when the Euler-Lagrange phase-field equations \cref{eq:strongform} derive from a single energy functional. For example, the variational consistency would be broken if the damage-evolution equation (\ref{eq:strongform}b) is modified without appropriately modifying the stress-equilibrium equation (\ref{eq:strongform}a) as well. Such variational 'breaking' is frequently done, for example, to introduce a distinct mode-II fracture toughness ($G_c^I,G_c^{II}$) \cite{zhang_modification_2017}. In such a case, the solution to the Euler-Lagrange equations \cref{eq:strongform} no longer minimises the energy functional \cref{eq:energy_regularized}.     

\subsection{Phase-field constitutive models}

Various constitutive modelling choices can be made to simulate brittle fracturing. In this work, the most commonly-employed models in the phase-field literature are studied;  the $AT1$ and $AT2$ models. Such models are proven to converge to a Griffith crack for a sufficiently small lengthscale  \cite{ambrosio_approximation_1990, wu_chapter_2020,de_lorenzis_numerical_2020}. For such models, the degradation of material stiffness is modelled using a quadratic degradation function, while the local dissipation function is linear (AT1) or quadratic (AT2):  
\begin{subequations}
    \begin{align}
    &\text{AT1: } \qquad {g(c)=(1-c)^2}, \qquad {\omega(c)=c} \qquad \implies \qquad c_\omega =\frac{2}{3}, \qquad w_1 = \frac{3\pazocal{G}_c}{8l}  \,,\label{eq:AT1} \\
    &\text{AT2: } \qquad {g(c)=(1-c)^2}, \qquad {\omega(c)=c^2} \quad\,\,\,\, \implies \qquad c_\omega =\frac{1}{2}, \qquad w_1 = \frac{\pazocal{G}_c}{2l}. \label{eq:AT2}
\end{align}
\end{subequations}

The AT1 model reproduces a more brittle response compared to the AT2 model, which does not have an elastic phase \cite{kristensen_assessment_2021}, and therefore is deemed more suitable to simulate failure in rocks. An important consequence of using such formulations is that the tensile strength of the material is dependent on the lengthscale parameter \cite{marigo_overview_2016, bilgen2020detailed}, i.e., 
\begin{equation}\label{eq:tensile_strength}
    \sigma^{AT1}_t = \sqrt{\frac{3EG_c}{8l}}\,, \qquad \sigma^{AT2}_t = \frac{3}{16}\sqrt{\frac{3EG_c}{l}}\,.
\end{equation}
Alternative constitutive models eliminate such a dependency, such as the cohesive zone method \cite{hansen-dorr_phase-field_2019,wu_length_2018}, but these are not studied herein, although they warrant further investigation. 
\hfill\newline

A choice for the elastic energy densities $\psi_r, \psi_d$ must also be made to determine the fracturing behaviour under compression. Various formulations have been proposed \cite{ambati_review_2015,amor_regularized_2009, miehe_phase_2010, zhou_phase-field_2019,de_lorenzis_nucleation_2022}. As the jointing within sedimentary layers occurs primarily under extension, it may seem that fracturing under compression is not of concern herein. Instead, it is common for displacement discontinuities, like cavities and fractures, to cause the background stress field to reverse in localised areas \cite{jaeger_fundamentals_2007,de_joussineau_can_2007,bai_closely_2000}, leading to locally compressive stresses within a regionally tensile stress-field, and \textit{vice-versa}. In this study, the volumetric-deviatoric split is tested, since it is well-studied \cite{amor_regularized_2009,ambati_review_2015}, and shares an intuitive similarity to elastic-visco-plastic modelling approaches \cite{gerya_introduction_2019}. The split is defined as
\begin{align}\label{eq:vol_dev_split}
    \psi^{d}_e(\uv) &= \frac{1}{2}K[\text{tr}^{+}(\boldsymbol{\varepsilon})]^2 + \mu\boldsymbol{\varepsilon}_D : \boldsymbol{\varepsilon}_D \,, \\
    \psi^{r}_e(\uv) &= \frac{1}{2}K[\text{tr}^{-}(\boldsymbol{\varepsilon})]^2 \, , 
\end{align}
where \(\text{tr}^{\pm}(\boldsymbol{\varepsilon}) = \text{max}(\pm \text{tr}(\boldsymbol{\varepsilon}), 0) \) and the double dot product of two matrices $A : B = \text{Trace}(AB^T)$. As a consequence of the definition of the stress \cref{eq:stress_def}, this gives 
\begin{equation}\label{eq:stress_vol_dev}
    \sigma_{ij} = g(c) \big( H^+(\varepsilon_{kk}) K \varepsilon_{kk}\delta_{ij} +  2\mu\varepsilon'_{ij}\big) 
                + H^-(\varepsilon_{kk}) K \varepsilon_{kk}\delta_{ij} \,,
\end{equation}
with the standard definition of the Kronecker delta $\delta_{ij}$ and the deviatoric strain $\varepsilon'_{ij} = \varepsilon_{ij} - \frac{1}{d}\varepsilon_{kk}\delta_{ij}$, with $d$ equal to the number of spatial dimensions. The Heaviside functions $H^+(x) = \boldsymbol{1}_{x\geq0}$ and $H^-(x)=\boldsymbol{1}_{x<0}$ are used to define when the material is in compression or tension. Consequently, Equation~\eqref{eq:stress_vol_dev} illustrates how both the bulk ($K$) and shear $(\mu)$ modulus of the rock degrade to zero for a fully damaged material under tension ($H^+(\varepsilon_{kk})=1$). Instead, when the material is under compression ($H^-(\varepsilon_{kk})=1$),  the bulk modulus is fully recovered, therefore preventing the un-physical interpenetration of damaged material. On the other-hand, \cref{eq:stress_vol_dev} shows how the shear modulus is not recovered upon compression, giving the possibility for the material to further fracture under `shear'. At this point, the link between damage models and the phase-field method can be noted, since the phase-field $c$ acts like a damage parameter that decreases the value of the elastic parameters $K$ and $\mu$ \cite{marigo_overview_2016}. This becomes apparent in the definition of the material stiffness tensor $C_{ijkl} = \frac{\partial\sigma_{ij}}{\partial\varepsilon_{kl}}$ 
\begin{equation}\label{eq:elasticity_tensor_C}
    C_{ijkl} = \bigg( \,\big[g(c)H^+(\varepsilon_{kk} ) + H^-(\varepsilon_{kk} )\big]K  - g(c)\frac{2}{d} \mu \bigg) \delta_{ij}\delta_{kl} + g(c) \mu (\delta_{ik}\delta_{jl} + \delta_{il}\delta_{kj} ) \, ,
\end{equation}
which now depends on the phase-field (damage) variable $c$. 
\hfill\newline

For the later visualisation of the results, it will be beneficial to define the following volumetric and deviatoric components of the energy density
\begin{equation}
    \psi_{vol} = g(c)\frac{1}{2}K[\text{tr}^{+}(\boldsymbol{\varepsilon})]^2 + \frac{1}{2}K[\text{tr}^{-}(\boldsymbol{\varepsilon})]^2\,, \qquad 
    \psi_{dev} = g(c) \mu\boldsymbol{\varepsilon}_D : \boldsymbol{\varepsilon}_D \,, 
\end{equation}
while the elastic energy density $\psi_e = \psi_{vol} + \psi_{dev}$, and the fracture energy density $\psi_f = G_c\gamma$.

\subsection{Finite element solution with monolithic trust region solver}
Since the governing equations \cref{eq:strongform} are variationally consistent, their numerical solution can be found by minimising the global energy functional $\pazocal{E}$, given by \cref{eq:energy_regularized}. 
In the literature, several monolithic solvers were proposed; see, for example~\cite{kopanivcakova2023nonlinear,farrell2017linear,lampron2021efficient}.
Here, we employ the trust region method, which has been shown to be effective in the context of phase-field fracture problems in~\cite{kopanicakova_recursive_2020}.
The trust region method is a globally convergent optimization method that utilizes gradient and Hessian information to minimize a function~\cite{nocedal_numerical_1999,Conn2000trust}. The finite element discretization of the energy functional, the Euler-Lagrange equations, and the Hessian detailed in \ref{sec:FE_discretisation}, along with a brief outline of the trust region algorithm. The code is open source and can be found in the following repository: \href{https://github.com/zulianp/utopia/tree/franetg}{https://github.com/zulianp/utopia/tree/franetg}. 
\hfill\newline

Here, we point out that the trust region method considered in this work is based on energy minimization.
This can be limiting in the case of the phase-field fracture models which do not maintain the variational consistency between the energy minimization problem and the utilized Euler-Lagrange equations. For example, several recently proposed phase-field fracture models are constructed by modifying the Euler-Lagrange equations with the aim to introduce a separate mode II fracture toughness for rocks \cite{zhang_modification_2017}. 
In this particular case, the implementation of the TR region method has to be adjusted such that it minimizes an appropriate merit function, e.g., $ \frac{1}{2} \| F(\hat{\uv}, \hat{\boldsymbol{c}})  \|^2$, where $F(\hat{\uv}, \hat{\boldsymbol{c}})$ denotes the discretized Euler-Lagrange equations.

\subsection{Utopia implementation}
We implemented the models, discretisation, solution algorithms, and post-processing tools utilising the open-source C++ nonlinear algebra library Utopia~\cite{zulian_large_2021} in combination with its PETSc~\cite{balay2020petsc} backend.
The PETSc structure-grid data manager (DM) is used for managing the discrete fields of phase and displacement, and we added the finite element discretisation of the phase-field model.
The incremental loading and nonlinear solution algorithms are realised with Utopia. 
We solve the trust region sub-problem with the Lanczos CG method~\cite{gould1999solving}, preconditioned using the Hypre's Boomer algebraic multigrid~\cite{falgout2002hypre}.

\subsection{Model setup: the reference case}

\begin{figure}[h]
\centering
\scalebox{0.8}{
\begin{tikzpicture}[scale=0.7]
    \pgfmathsetmacro{\height}{5} \pgfmathsetmacro{\length}{16} \pgfmathsetmacro{\heightcm}{50} \pgfmathsetmacro{\lengthcm}{160} \pgfmathsetmacro{\blay}{2} \pgfmathsetmacro{\tlay}{1.0} \pgfmathsetmacro{\rad}{0.3} 

      \draw[color=black, very thick, fill=gray!10!] (0.0, 0.0) rectangle (\length, \height);
      \draw[color=gray, very thin, fill=gray!50] (0.04, \blay) rectangle (\length-0.04, \blay+\tlay);

\draw[line width=0.75mm, color=black] (\length+0.2, 0.1) -- (\length+0.2, \height-0.1);

\draw[tipC-tipA, fill=black, line width=0.3mm](\length+0.4,\height/2) -- node[above,yshift=0.5cm]
      {\footnotesize\Large$\qquad u_x=Ut$}  (\length+2,\height/2);  

\draw[tipA-tipA, fill=black, thin](0,-1.5) -- node[below]
      {\footnotesize\large $L=\lengthcm\,$cm}  (\length,-1.5);  \draw[tipA-tipA, fill=black, thin](-1.5,0) -- node[rotate=90, above]
      {\footnotesize\large$H=\heightcm\,$cm}  (-1.5,\height);  

\draw[tipB-tipB, fill=black, thin](\length-4,0.0) -- node[right]
      {\footnotesize\large $H_s=22\,$cm}  ( \length- 4, \blay ); \draw[tipB-tipB, fill=black, thin](\length-4,\blay) -- node[right]
      {\footnotesize\large $H_d=6\,$cm }  ( \length- 4, \blay+\tlay ); \draw[tipB-tipB, fill=black, thin](\length-4 ,\blay+\tlay) -- node[right]{\footnotesize\large $H_s=22\,$cm}  ( \length-4, \height ); 

      \node[font=\bfseries\large, right] at (0.5, \height/2) 
      { Dolostone layer: $E_d,\, \nu_d,\, G^d_c$ };
      \node[font=\bfseries\large, right] at (0.5, \blay/2) { Shale layer: $E_s,\, \nu_s,\, G^s_c$ };
      \node[font=\bfseries\large,right] at (0.5, \blay+\tlay+\blay/2) 
      { Shale layer: $E_s,\, \nu_s,\, G^s_c$ };

\draw[thin] \foreach \y in {0.75,2,..., \height} { (-\rad, \y) circle (\rad cm)};
        \draw[line width=0.5mm, color=black, thin] (-\rad-\rad, 0.4) -- (-\rad-\rad, \height);

\draw[thin] \foreach \x in {1, 3, ..., \length} { (\x,-\rad) circle (\rad cm)};
        \draw[line width=0.5mm, color=black, thin] (0.5, -\rad-\rad) -- (\length, -\rad-\rad);

\node[regular polygon, regular polygon sides=3, fill=gray!10, draw=black, minimum size = 0.9cm, scale=0.6, anchor = north] at (0, 0) {};
         \node[regular polygon, regular polygon sides=3, fill=gray!10, draw=black, minimum size = 0.9cm, scale=0.6,rotate=-90, anchor = north] at (0 , 0) {};

\draw[thin] \foreach \x [evaluate=\x as \y using \x/7 ]   in {1,...,5} { (\y-0.42, -\rad-\rad) -- (\y-0.42 -0.15, -\rad-\rad-0.25) };
        \draw[thin, rotate=-90] \foreach \x [evaluate=\x as \y using \x/7 ]   in {1,...,5} { (\y-0.42, -\rad-\rad) -- (\y-0.42 -0.15, -\rad-\rad-0.25) };
\end{tikzpicture}  }
\label{fig:boundary_conditions}
\caption{The reference case: A three-layer sedimentary sequence composed of a stiff but weak dolostone layer surrounded by more compliant but strong shale layers. Displacements in the horizontal (x) dimension are imposed on the right boundary at each time step, while the bottom and left boundaries are allowed to slide.}
\end{figure}
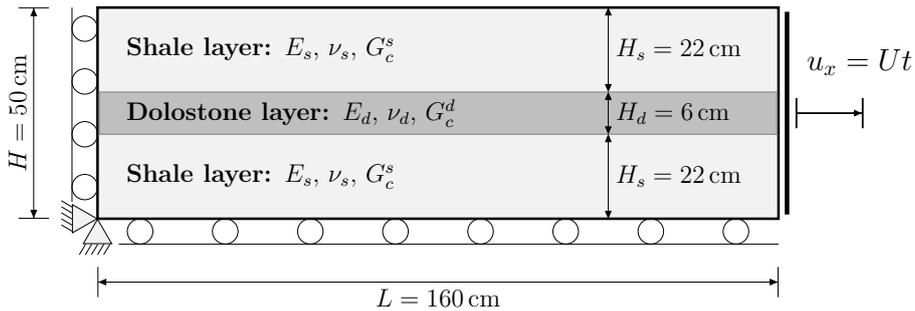

As outlined in the introduction, this work does not aim to provide an exhaustive sensitivity study of jointing, but rather, is geared towards highlighting the dominant phase-field behaviours that determine the capability of the method. During this work, it was found that a significant part of the understanding can be conveyed with a small range of simulations and model parameters. Consequently, it is useful to define a reference case, and any deviations from such a reference case will be explicitly mentioned.  
\hfill\newline

\begin{table}
    \centering
    \title{Mechanical parameters for Three-layer model}
    \begin{tabular}{c|cccccc|cc }
&  \multicolumn{6}{c|}{\textbf{Mechanical Parameters}} & \multicolumn{2}{c}{\textbf{Numerical Parameters}} \\ \hline
    Layer  & Height & E & $\nu$ & $\pazocal{G}_c$  & $l$ & $\sigma_t$ & $\varsigma_{irr}$ & $\varsigma_{bou}$\\ \hline
        Dolostone  & 6 cm & 50 GPa & 0.25 & 20 J/m & 1.5 cm & $5$ MPa & 0.02 & 0.02 \\
        Shale & 22 cm  & 10 GPa & 0.35 & 160 J/m & 1.5 cm & $6.3$ MPa & 0.02 & 0.02 \\ 
\end{tabular}
    \caption{Values of the mechanical and numerical parameters chosen for the dolostone and shale layers in the reference case of the Three-layer model. Tensile strength $\sigma_t$ is not prescribed, but derived from \cref{eq:tensile_strength}.  }
    \label{tab:mech_params}
\end{table}

The model dimensions and mechanical parameters for the reference case are summarised in \cref{fig:boundary_conditions} and \cref{tab:mech_params}, respectively. The model setup consists of a three-layered sedimentary sequence that contains a stiff dolostone layer with height $H_d=6\,$cm surrounded by more compliant shale layers with height $H_s=22\,$cm. With the ratio $H_s/H_d\sim3.7$, boundary effects have been shown to be limited \cite{chukwudozie_new_2013}. The dolostone layer is modelled as both stiffer and \textit{weaker} than the shale layers, by using a higher Young's modulus and a lower fracture toughness and tensile strength compared to the shale. The parameters are chosen similar to Tang \textit{et.al}~\cite{tang_fracture_2008} in order to aid comparison. The total length and height of the model are $L=160\,$cm and $H=50\,$cm, respectively. A layer-parallel extension is applied on the model by imposing a horizontal ($x$) displacement that increases in time ($t$) according to $u_x = U t$, with $U=0.8\,$cm/s. Larger time-steps of $\Delta t=1e^{-4}\,$s were chosen at the start of the model, and then decreased to $\Delta t=1e^{-5}\,$s close to the point when the first fracture nucleates. The displacements on the bottom and left boundaries are fixed in the normal direction and are free to move tangential to the boundary, while the displacement in the bottom left corner is fixed in both the $x$ and $y$ dimension. A uniform mesh with resolution $h=0.2\,$cm was chosen, which is sufficiently fine compared to the lengthscale $l=1.5\,$cm. 
To enforce irreversibility and boundedness of the phase-field, we use the penalty method~\cite{gerasimov_penalization_2019}.
More details regarding the implementation can be found in the \ref{sec:FE_discretisation}.

   \section{Jointing in sedimentary sequences: reference model results}
\label{sec:pf_perspective}

\begin{figure}
\centering
  \includegraphics[width=0.9\linewidth]{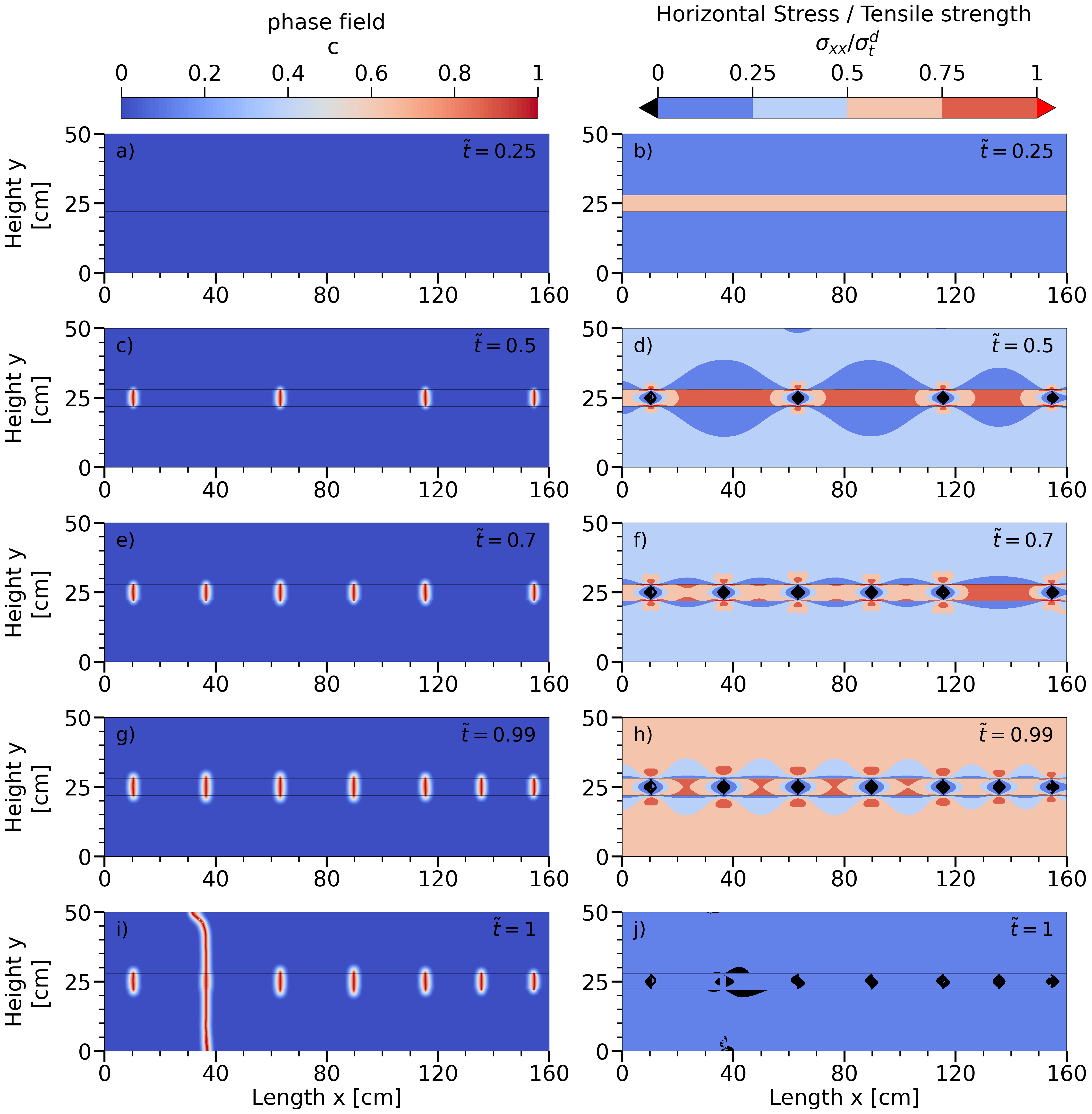}
  \caption{Results of the AT1 model for the phase-field $c$ (left) and the horizontal stress $\sigma_{xx}$ (right) normalised by the tensile strength of the dolostone layer $\sigma_t=5\,$MPa. The geometries and parameters used in the simulation are shown in \cref{tab:mech_params}. Various snapshots are shown at dimensionless times $\tilde{t}$, normalised by the time at failure $t_{fail}=0.00556\,$s,  at an applied boundary displacement of approximately $u_x=0.44 mm$. }
  \label{fig:ThreeLayer_60}
\end{figure}

The reference model simulation shows an overall intuitive behavior of the system from initial stress build-up to a final complete failure. This is defined here as formation of a fracture through the whole vertical extent of the specimen.
Key steps of the process are displayed in \cref{fig:ThreeLayer_60} for the AT1 model, with model parameters and boundary conditions summarised in \cref{tab:mech_params} and \cref{fig:boundary_conditions}. The dimensionless time $\tilde{t}$ is defined by normalising by the time to complete failure. Initially, the horizontal stress in the model builds up faster in the stiffer dolostone layer (\cref{fig:ThreeLayer_60}b), resulting in the nucleation of the first vertical joints, which can be noted by the phase-field approaching the fully damaged state $c=1$ at four locations (\cref{fig:ThreeLayer_60}c). At this point, the horizontal stress is most tensile away from the joints, reaching its largest value in the middle between two joints (\cref{fig:ThreeLayer_60}d). As the specimen is subject to increasing horizontal strain over time, additional joints infill between the existing joints, coinciding with the locations of highest stress (\cref{fig:ThreeLayer_60}e-h). The results at dimensionless time $\tilde{t}\simeq0.7$ highlights the influence the spacing of joints has on the horizontal stress of the middle layer; stress levels between closer spaced joints ($\sim30\,cm$) fail to reach more than $75\%$ of the tensile strength of the dolostone, while stress levels between joints with larger spacing ($\sim40\,cm$) reach higher levels. The last two time snapshots (\cref{fig:ThreeLayer_60}g,h-i,j) represent the scenario before and after complete failure of the specimen. In such a case, a level of joint saturation has been reached, with most levels of stress within the layer below $75\%$ of the dolostone's tensile strength ($\sigma^d_t=5\,$MPa). Although not apparent in the figure, the stress in the middle dolostone layer \textit{decreases} during the final time-steps before complete failure. Instead, large  horizontal stresses can be seen to develop at $\tilde{t}=0.99$ within the bounding shale layers, above and below the fracture tips. This coincides with an incremental propagation of the phase-field at the fracture tips, eventually resulting in the instantaneous failure of the specimen at $\tilde{t}=1$. Upon complete failure, the horizontal stresses do not drop completely to zero, but remain at a few hundred pascals, due to a numerical residual stiffness imposed on fully cracked specimens. 
\hfill\newline

\begin{figure}[h]
  \centering
  \includegraphics[width=.6\linewidth]{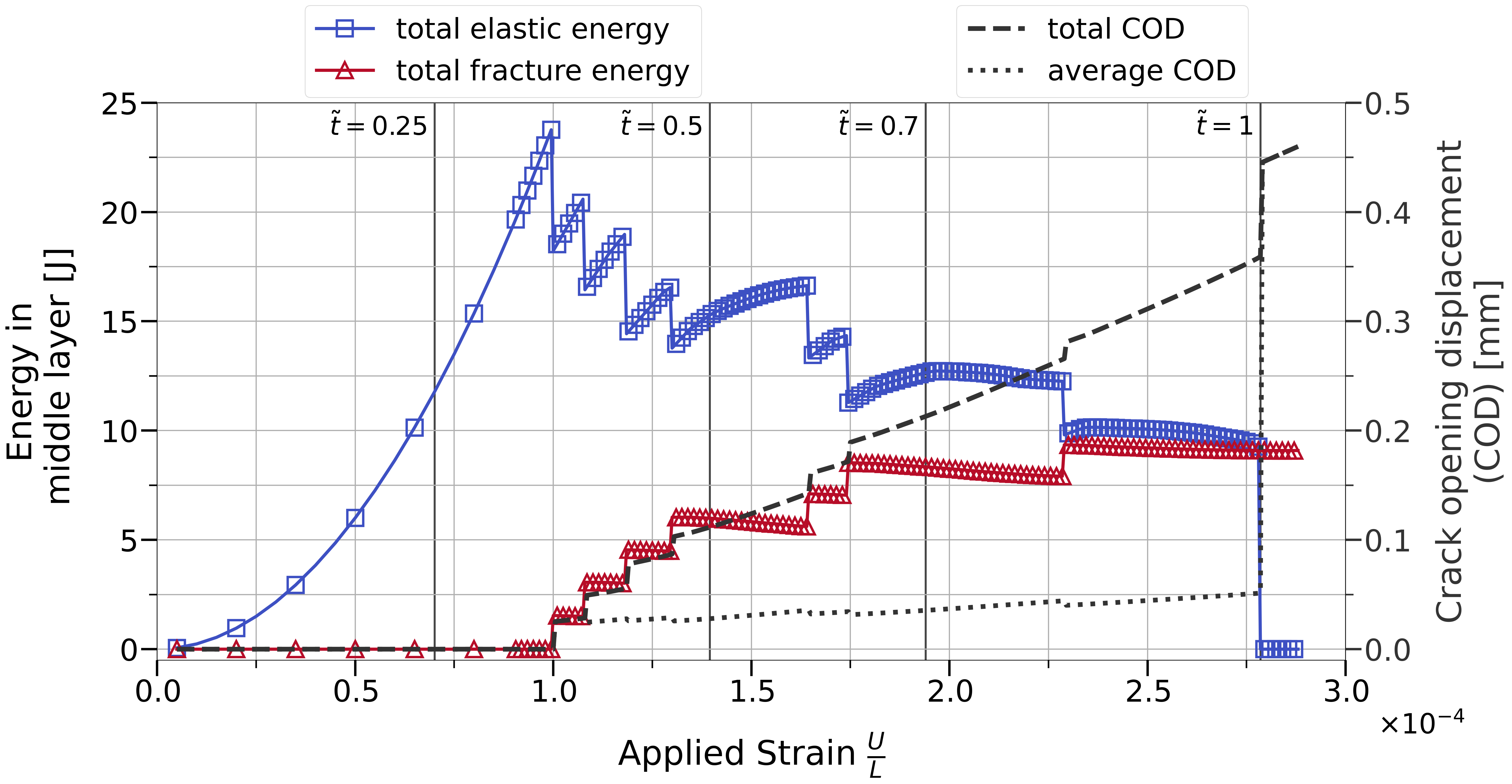}
  \caption{Evolution of the total elastic and fracture energy (left axis) alongside the total and average fracture crack opening displacement (right axis) within the middle dolostone layer for the AT1 model. The COD is calculated from the difference in horizontal displacement taken at the middle of the fracture, where it is expected to be greatest. The dimensionless times chosen for the phase-field snapshots in \cref{fig:ThreeLayer_60} are superimposed on the plot.}
  \label{fig:EnergyCOD_60}
\end{figure}

\subsection{Energy balance}

Further insight on the nucleation of joints can be obtained by looking at the global energy balance of the middle dolostone layer, shown in \cref{fig:EnergyCOD_60}. The energies plotted correspond to the total elastic and fracture energy in the middle layer. Initially, the elastic energy in the layer grows quadratically due to the linear increase in applied strain and stress. A peak of $24\,$J is reached for the total elastic energy in the layer, after which the first fracture nucleates, and the total elastic energy in the layer is `traded' for an increase in the fracture energy and a decrease in the external work done on the system, since the specimen is now more compliant. The elastic energy rises in between each fracturing event, but each subsequent fracturing drives the elastic energy lower overall with increasing time. A turning point occurs at an applied strain of $U/L \sim 1.4e^{-4}$, which coincides with the phase-field distribution at $\tilde{t}\sim0.5$ in \cref{fig:ThreeLayer_60}(c,d). At this point, the dilation of fractures begins to accommodate the additional strain, as is shown by the right axis of \cref{fig:EnergyCOD_60}, which shows a significant growth of the combined crack opening displacement (COD) of all fractures (measured at the midline of each fracture). There is a reduction in the rate of increase in elastic energy at such points in time (e.g $\tilde{t}\sim0.5, \, 0.7$). This is caused by a lower effective stiffness of the dolostone layer with increasing damage, and also because of the incremental propagation of the joints into the shale, thereby releasing stress in the dolostone layer. As the fractures continue to dilate, the stress intensities at the fracture tips increase to the point where propagation occurs within the surrounding shale layers ($U/L>2.5e^{-4}$), causing the rupture of the whole specimen. 
\hfill\newline

\begin{figure}[h]
\centering
  \includegraphics[width=.7\linewidth]{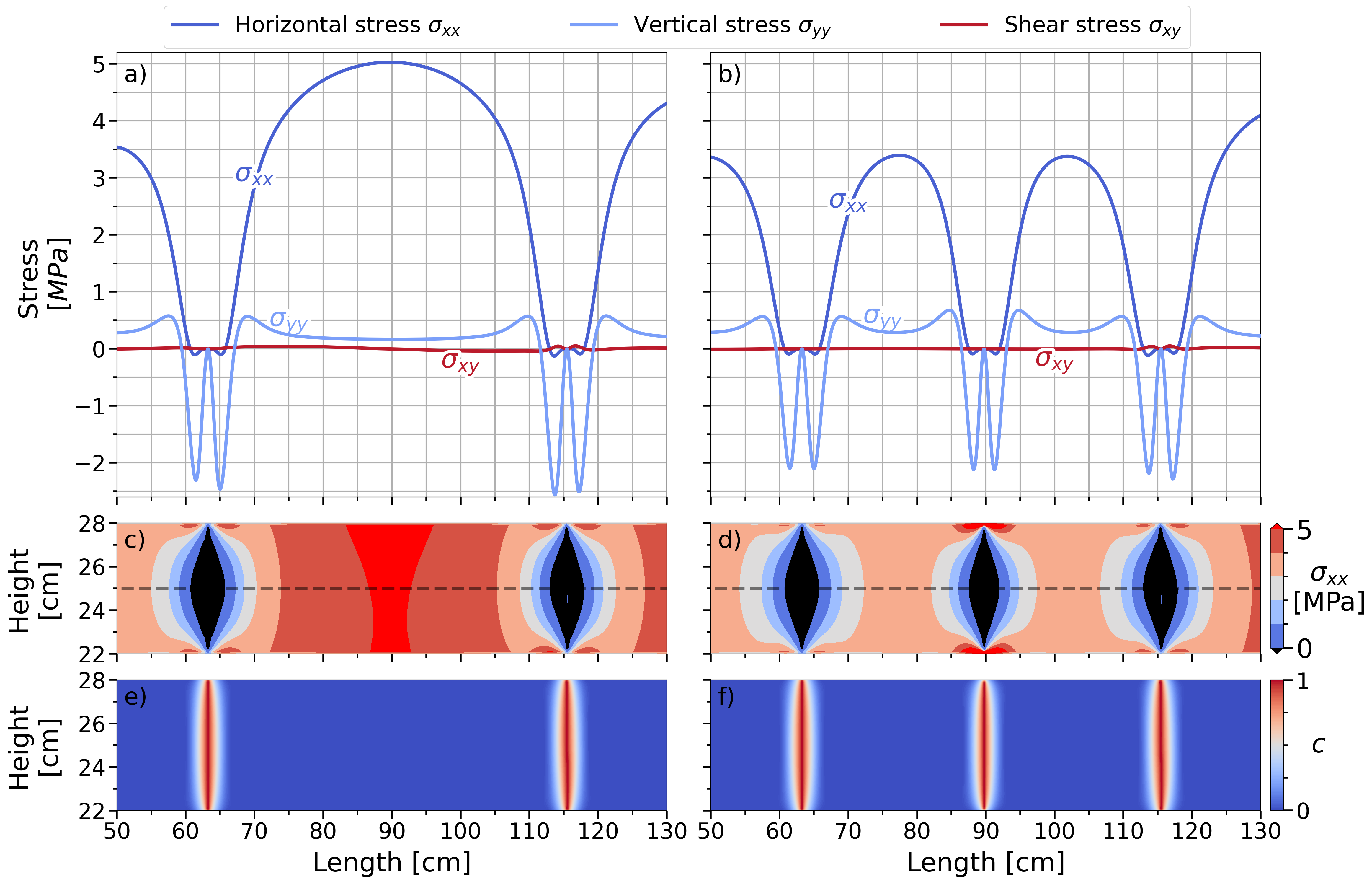}
  \caption{Distributions of stress (a,b,c,d) and phase-field (e,f) between two joints within the middle dolostone layer for two snapshots in time for the AT1 model. The first time snapshot (a,c,e) is the timestep before a third joint initiates in the middle between the two existing joints. Horizontal stress $\sigma_{xx}$ is plotted in (c,d) and shows a dashed horizontal line slice which shows the variation of various stresses in (a,b).}
  \label{fig:StressAnalysis_60}
\end{figure}

\subsection{Joint interaction \& stress shadow}
A closer look at the distribution of stress between two joints within the middle dolostone layer is shown in \cref{fig:StressAnalysis_60}. The left (a,c,e) and right (b,d,f) side of the plot marks the time step before and after an additional joint nucleates in the layer, respectively. This is noted by observing the phase-field distribution in \cref{fig:StressAnalysis_60}e-f. Compressive stresses can be seen to develop close to the fracture (black area in \cref{fig:StressAnalysis_60}c,d), and no stresses are present where the phase-field values reach $c=1$. \cref{fig:StressAnalysis_60}(a) shows the values of stress along a horizontal slice that passes through the middle of the dolostone layer, where compressive $\sigma_{yy}$ values surpass $2\,$MPa close to the fractures. Smaller compressive stresses are observed for the horizontal stress $\sigma_{xx}$, which follow a trend consistent with the results shown from Bai \cite{bai_explanation_2000}. Such a feature is known to occur due to the vertical shortening of the layer caused by the horizontal dilation of the fractures, resulting primarily in compressive vertical stresses $\sigma_{yy}$. The nucleation of additional joints at later times (\cref{fig:StressAnalysis_60}b) cause a decrease in $\sigma_{yy}$, since the horizontal strain is now accommodated by the additional third joint; reducing the dilation of each individual joint. Fracture nucleation occurs when the hortizontal stresses that reach the tensile strength $\sigma_{xx}=5\,$MPa span the entire height of the dolostone layer, as can be seen in \cref{fig:StressAnalysis_60}. Although \cref{fig:StressAnalysis_60} shows shear stresses $\sigma_{xy}$ are close to zero in \cref{fig:StressAnalysis_60}(a-b), their magnitude increases when approaching the top and bottom of the dolostone layer. 
\hfill\newline

\begin{figure}[h]
  \centering
  \includegraphics[width=.7\linewidth]{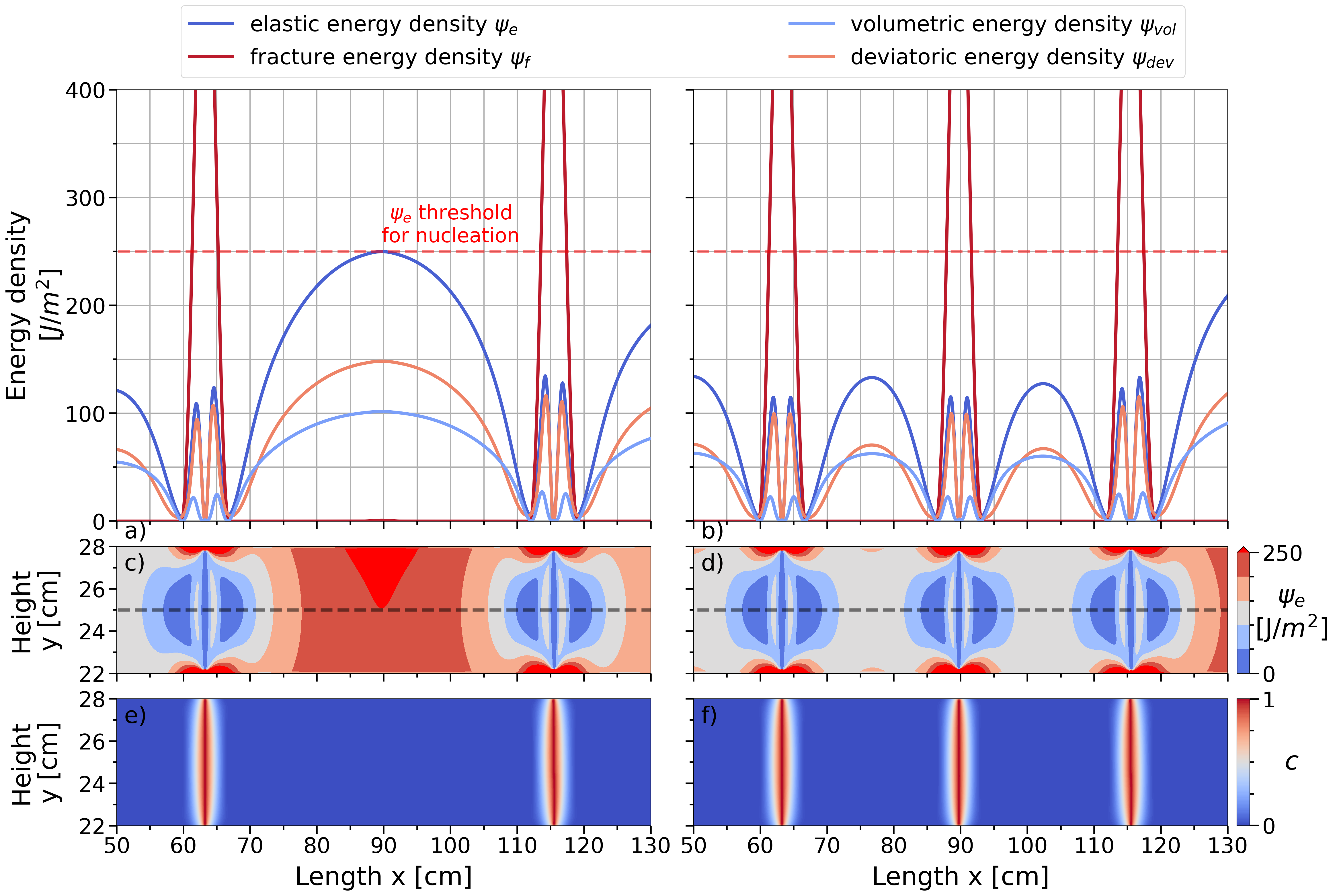}
  \caption{Distributions of energies (a,b,c,d) and phase-field (e,f) between two joints within the middle dolostone layer for two snapshots in time for the AT1 model. The first time snapshot (a,c,e) is the timestep before a third joint initiates in the middle between the two existing joints. Elastic energy density $\psi_{e}$ is plotted in (c,d), while profiles of volumetric, deviatoric, elastic and fracture energy are plotted (a,b) for the dashed horizontal line at the height of $y=25\,$cm, which corresponds to the midline of the layer.}
  \label{fig:EnergyAnalysis_60}
\end{figure}

To better understand the mechanisms of nucleation and propagation within phase-field models, it is helpful to observe the distribution and balance of energies in the dolostone layer. Analogous to \cref{fig:StressAnalysis_60}, \cref{fig:EnergyAnalysis_60} shows the elastic energy of a subsection of the dolostone layer (c,d), along with energy profiles along the layer's midline at a height of $y=25\,$cm (a,b). The left and right hand sides of \cref{fig:EnergyAnalysis_60} correspond to the same snapshots in time as in \cref{fig:StressAnalysis_60}, giving the same phase-field distribution in subfigures (e, f). Before the nucleation of the third joint, the distribution of elastic energy (\cref{fig:EnergyAnalysis_60}c) reaches $\psi_{e}=250\,$J/m$^2$ between two joints. The magnitude of $\psi_e$ is primarily controlled by the value of horizontal stress, which is equal to the tensile strength $\sigma_{xx}=\sigma_t$ at such a location. Roughly equal portions of the volumetric $\psi_{vol}$ and deviatoric $\psi_{dev}$ energy densities contribute to the total elastic energy density at the midpoint between two joints. After the nucleation of the third joint (\cref{fig:EnergyAnalysis_60}d), the elastic energies have decreased ($\psi_e, \psi_{vol},\psi_{dev}$) and are balanced by an increase in the fracture energy which reaches $\psi_f\sim800\,$J/m$^2$ at the location of the third joint at length $x=90\,$cm.  A more nuanced balance of energy occurs close to the fracture, where both fracture and elastic energies are present. In particular, \cref{fig:EnergyAnalysis_60}(b) show that small volumetric stress $\psi_{vol}<25\,$J/m$^2$, and larger deviatoric stresses $\psi_{dev}\sim100\,$J/m$^2$ persist for fracture energies $\psi_f\lesssim300\,$J/m$^2$. The peaks in $\psi_{vol},\psi_{dev}$ in such a region occurs at phase-field values of $c\sim0.33$, and corresponds with the location where volumetric strains and stresses have become negative. The consequence of choosing the volumetric-deviatoric energy split (\cref{eq:vol_dev_split}) in such a region leads to the degradation of the shear modulus $\mu$, while restoring the bulk modulus $K$ to its undamaged form. The large peaks in deviatoric energy will be shown in \cref{sec:pf_challenges} to lead to premature fracturing under-compression within the AT1 model. 
\hfill\newline

\subsection{Fracture spacing}
\begin{figure}[t]
  \begin{minipage}{0.47\linewidth}
    \includegraphics[width=\linewidth]{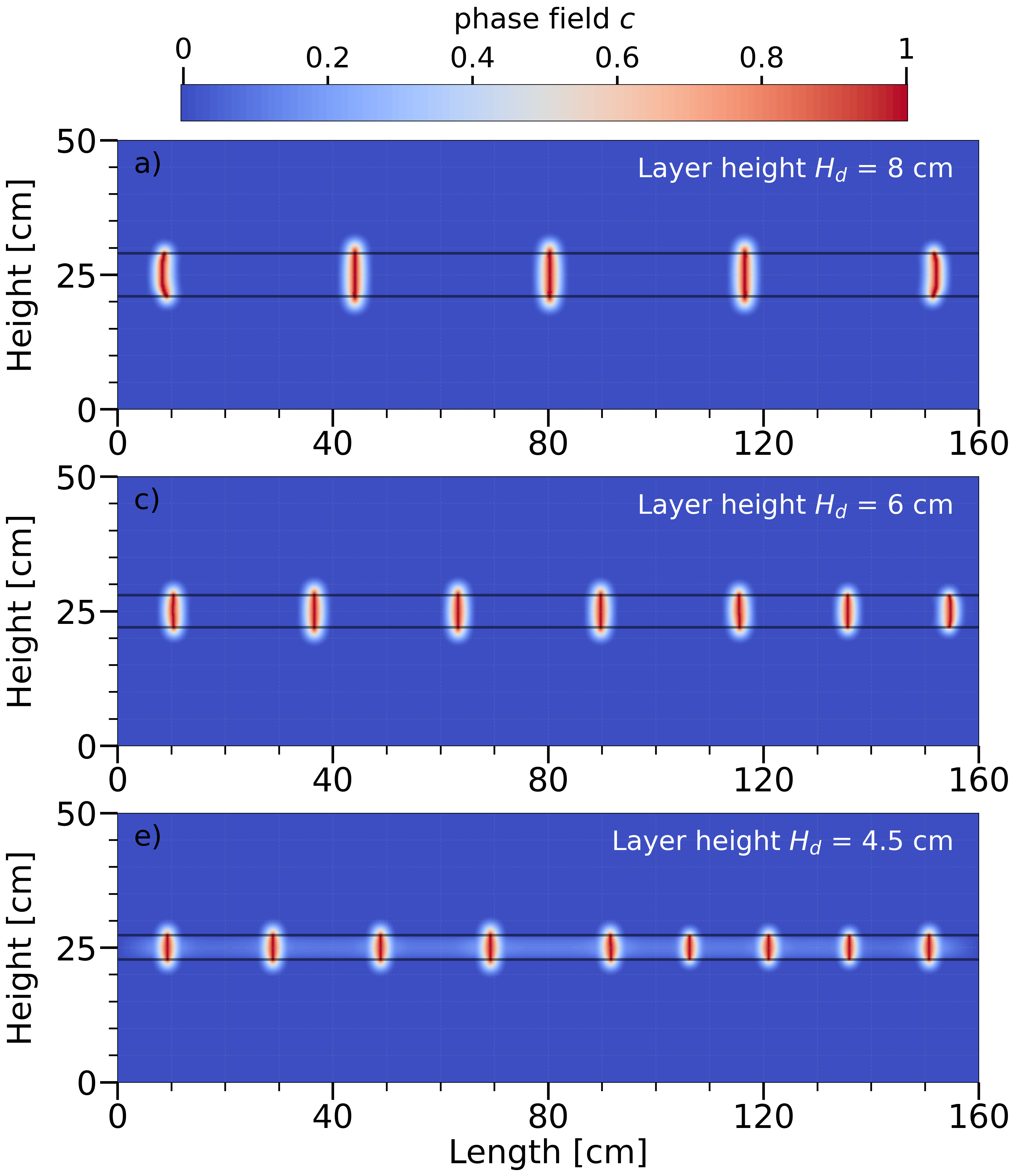}
  \end{minipage}
  \hfill
  \begin{minipage}{0.52\linewidth}
    \includegraphics[width=\linewidth]{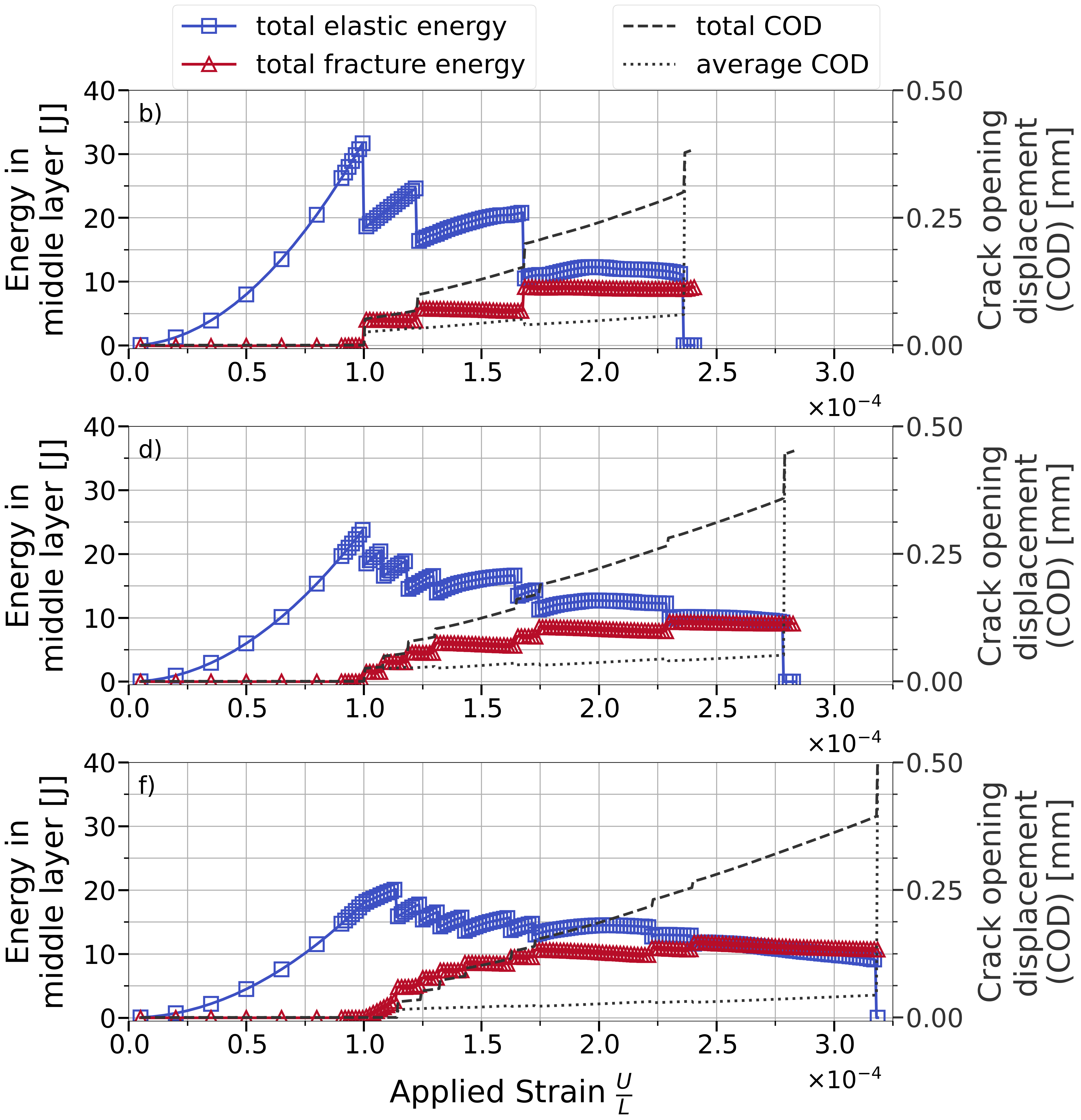}
  \end{minipage}
  \caption{Effect of middle layer height on fracture spacing in the AT1 model. Left: Phase-field distribution before complete failure of the sample. Right: Evolution of the total energy-strain in the middle dolostone layer, alongside crack opening displacement measured at the middle of the fracture. Mechanical parameters are the same as in \cref{tab:mech_params}}
  \label{fig:fracture_spacing_correlation}
\end{figure}

Outcrops display a positive correlation between the height of the dolostone layer and spacing between fractures. The phase-field method reproduces such a positive correlation, but overestimates fracture spacings by a factor of 4. This is shown in \cref{fig:fracture_spacing_correlation}(a,c,e), where the phase-field distributions for simulations with varying layer heights $H_d=4.5,6,8\,$cm are shown for the AT1 model, before the onset of total failure of the specimen. A clear correlation can be noted  whereby thinner dolostone layers achieve a smaller fracture spacing before total failure. However, the average fracture spacing for the increasing layer heights is approximately $17,\,20, \&\, 35\,$cm, giving fracture spacing to height ratios of $S/H_d\sim3.7,\,3.3,\, 4.4$, respectively. Such a feature can also be noted in \cite{chukwudozie_new_2013}, whereby the authors utilised the AT2 model.  Such ratios overestimate the majority of fracture spacings observed in outcrop studies that report ratios $S/H_d\lesssim1$ \cite{bai_fracture_2000}. This feature of phase-field models is further analysed in \cref{sec:pf_challenges}. 

Despite the discrepancy, the phase-field method offers further insights into the jointing process. By looking at the energy balance in \cref{fig:fracture_spacing_correlation}(b,d,f), thinner layers can be seen to sustain higher levels of strain before failing. This feature is explained by the larger extent of cumulative dilation (total COD) shared amongst joints in thinner layers, thereby increasing the effective compliance of the layer. Lower fracture spacing ratios $S/H_d$ lead to lower levels of dilation per joint, therefore delaying the propagation of the joint in the surrounding layers. For the thinnest layer ($H_d\leq4.5\,$cm), a diffuse damage can be seen to develop by observing the rise in fracture energy (\cref{fig:fracture_spacing_correlation}f) before the nucleation of the first fracture. Such a feature is a peculiarity inherent to the AT1 and AT2 models, arising when the phase-field lengthscale $l$ is comparable to the dimensions of the layer. This is due to the stability of homogeneous states that delay fracture nucleation at the expense of an increasing uniform (homogeneous) damage field (\cite{marigo_overview_2016}). The diffuse damage ($c\sim0.1$) in the thinnest layer further increases the effective compliance of the layer, allowing larger strains before complete failure of the specimen. In reality, whether such diffuse damage occurs in thin interbedded sequences is yet to be determined.  
\hfill\newline

   \section{Phase-field challenges in reproducing jointing}
\label{sec:pf_challenges}

\subsection{The distortion of stress by the phase-field lengthscale}\label{sec:stress_distorting_lengthscale}

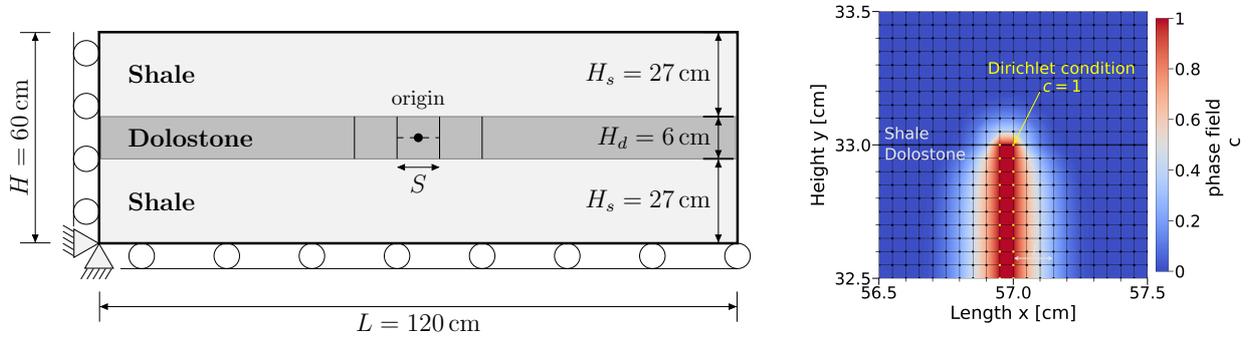
\begin{figure}
\centering
\begin{subfigure}{0.6\textwidth}
\scalebox{0.8}{
\begin{tikzpicture}[scale=0.7]
    \pgfmathsetmacro{\height}{5} \pgfmathsetmacro{\length}{15} \pgfmathsetmacro{\heightcm}{60} \pgfmathsetmacro{\lengthcm}{120} \pgfmathsetmacro{\blay}{2} \pgfmathsetmacro{\tlay}{1.0} \pgfmathsetmacro{\rad}{0.3} 

      \draw[color=black, very thick, fill=gray!10!] (0.0, 0.0) rectangle (\length, \height);
      \draw[color=gray, very thin, fill=gray!50] (0.04, \blay) rectangle (\length-0.04, \blay+\tlay);

\draw[tipA-tipA, fill=black, thin](0,-1.5) -- node[below]
      {\footnotesize\large $L=\lengthcm\,$cm}  (\length,-1.5);  \draw[tipA-tipA, fill=black, thin](-1.5,0) -- node[rotate=90, above]
      {\footnotesize\large$H=\heightcm\,$cm}  (-1.5,\height);  

\draw[tipA-tipA, fill=black, thin](\length-0.45,0.0) -- node[left]
      {\footnotesize\large $H_s=27\,$cm}  ( \length-0.45, \blay ); \draw[tipA-tipA, fill=black, thin](\length-0.45,\blay) -- node[left]
      {\footnotesize\large $H_d=6\,$cm }  ( \length-0.45, \blay+\tlay ); \draw[tipA-tipA, fill=black, thin](\length-0.45 ,\blay+\tlay) -- node[left]{\footnotesize\large $H_s=27\,$cm}  ( \length-0.45, \height );

      \node[font=\bfseries\large, right] at (0.5, \height/2) 
      { Dolostone};
      \node[font=\bfseries\large, right] at (0.5, \blay/2) { Shale };
      \node[font=\bfseries\large,right] at (0.5, \blay+\tlay+\blay/2) 
      { Shale };

\draw[thin] \foreach \y in {0.75,2,..., \height} { (-\rad, \y) circle (\rad cm)};
        \draw[line width=0.5mm, color=black, thin] (-\rad-\rad, 0.4) -- (-\rad-\rad, \height);

\draw[thin] \foreach \x in {1, 3, ..., \length} { (\x,-\rad) circle (\rad cm)};
        \draw[line width=0.5mm, color=black, thin] (0.5, -\rad-\rad) -- (\length, -\rad-\rad);

\node[regular polygon, regular polygon sides=3, fill=gray!10, draw=black, minimum size = 0.9cm, scale=0.6, anchor = north] at (0, 0) {};
         \node[regular polygon, regular polygon sides=3, fill=gray!10, draw=black, minimum size = 0.9cm, scale=0.6,rotate=-90, anchor = north] at (0 , 0) {};

\draw[thin] \foreach \x [evaluate=\x as \y using \x/7 ]   in {1,...,5} { (\y-0.42, -\rad-\rad) -- (\y-0.42 -0.15, -\rad-\rad-0.25) };
        \draw[thin, rotate=-90] \foreach \x [evaluate=\x as \y using \x/7 ]   in {1,...,5} { (\y-0.42, -\rad-\rad) -- (\y-0.42 -0.15, -\rad-\rad-0.25) };

\draw (\length/2 - \tlay/2, \blay) -- (\length/2 - \tlay/2, \blay+\tlay);
    \draw (\length/2 + \tlay/2, \blay) -- (\length/2 + \tlay/2, \blay+\tlay);
    \draw (\length/2 - 3 * \tlay/2, \blay) -- (\length/2 - 3 * \tlay/2, \blay+\tlay);
    \draw (\length/2 + 3 * \tlay/2, \blay) -- (\length/2 + 3 * \tlay/2, \blay+\tlay);

    \draw[tipB-tipB, fill=black, thin](\length/2 - \tlay/2,\blay-0.2) -- node[below]{\footnotesize\large $S$}  ( \length/2 + \tlay/2, \blay-0.2 ); 

    \draw[dashed] (\length/2 - \tlay/2, \blay+\tlay/2) -- (\length/2 + \tlay/2, \blay+\tlay/2);

    \fill[black] (\length/2, \height/2) circle (3pt) node [above,yshift=10pt]{origin};

\end{tikzpicture}  }
\label{fig:cracked_setup}
\end{subfigure}
\hfill
\begin{subfigure}{0.35\textwidth}
     \raisebox{4ex}{\includegraphics[width=\textwidth]{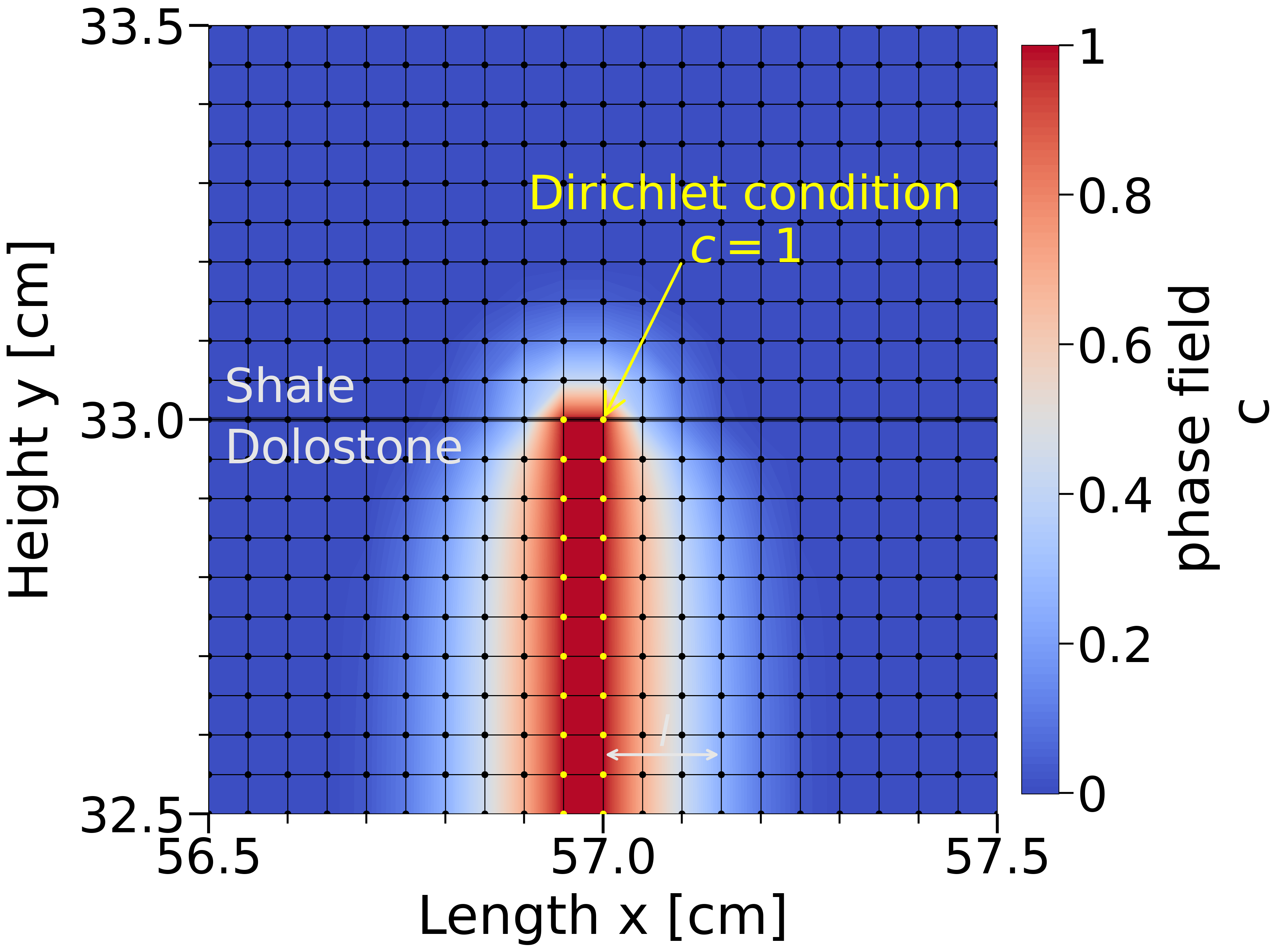}}
\end{subfigure}
\caption{(Left) Boundary condition and geometry of $4$ regularly spaced joints prescribed within a stiff dolostone layer. The central black dot denotes the origin of the stress plot shown in \cref{fig:StressTransitionAT1}. (Right) The initialisation of the phase-field for one of the joints. The fracture toughness of the shale is $G_c^s=200$J/m and the poisson ratio is the same for all layers $\nu=0.25$, while other parameters are kept as in \cref{tab:mech_params}.}
\end{figure}

\begin{figure}
\centering
  \includegraphics[width=0.9\linewidth]{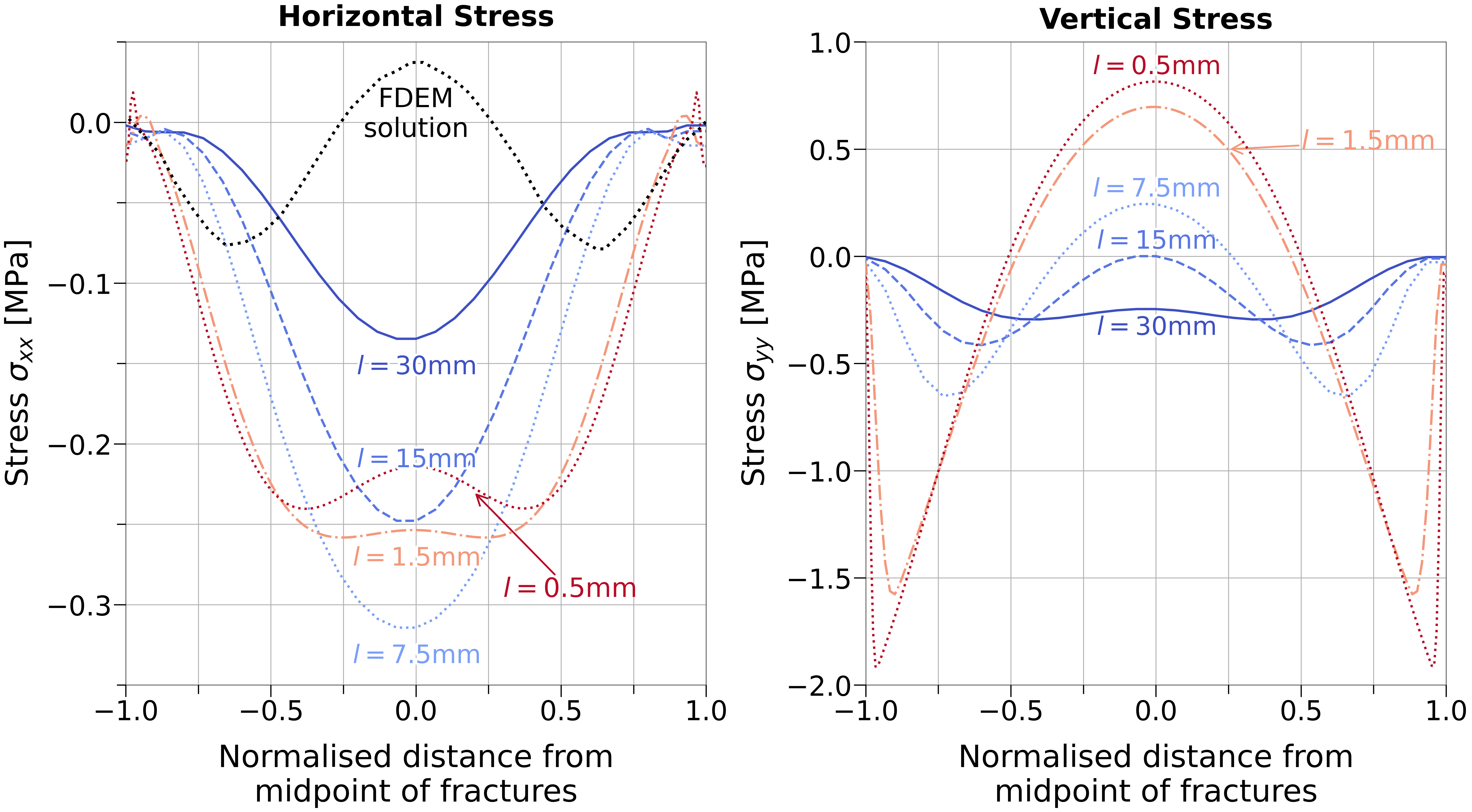}
  \caption{Variation of stress between joints at various phase-field lengthscales. The $x$ axis has origin as shown in \cref{fig:cracked_setup}, with values of $-1$, $1$ denoting the location of the central left and right joints, respectively. At very small lengthscales, horizontal stresses start to converge towards the FDEM solution of Bai and Pollard \cite{bai_fracture_2000} (Fig 2, $S/T_f=1$), but do not reach it.   }
  \label{fig:StressTransitionAT1}
\end{figure}

In the following section, we explore the significance of the phase-field lengthscale in reproducing the stress-distribution between fractures. Sharp-crack models, like the FDEM approach of Bai and Pollard \cite{bai_fracture_2000}, propose a stress-transition theory to explain the spacing of joints in sedimentary layers. The stress-transition theory rests on the observation that layer-parallel stresses cease to be solely tensile once joint spacings are smaller than layer heights.  Instead, compressive stresses exist between joints with smaller joint spacings, suppressing the infilling of additional fractures as a result. As a consequence, the stress-transition theory predicts fracture spacings equal to layer heights $S/H_d=1$, which coincides with the onset of compressive stresses. This was shown to be independent to the level of applied strain and weakly dependent on contrasts in Young's modulus and poisson ratio between layers. The stress-transition theory does not take into account any layer-parallel shearing or interface effects, and yet, reconciles a range of geological observations of jointing \cite{bai_closely_2000,bai_fracture_2000}. 

To evaluate the phase-field method's ability in predicting the stress transition, a geometrical setup akin to Bai and Pollard is used, which is shown in \cref{fig:cracked_setup}. Four joints are prescribed with uniform spacing equal to the layer height $S=H_d$. Simulations for various lengthscales are conducted, which are stopped after an applied strain of $\varepsilon=0.0001$ is reached. The FDEM solution of Bai $\&$ Pollard (fig 2, $S/T_f=1$) is scaled by the applied strain and young's modulus and plotted in \cref{fig:StressTransitionAT1} to enable a comparison against the phase-field simulations herein. Only results for the AT1 model are shown in \cref{fig:StressTransitionAT1}, as results for the AT2 model where found to be similar. 

Prohibitively small lengthscales appear to be required to reproduce the stress field predicted by FDEM modelling. The results for the horizontal stress in \cref{fig:StressTransitionAT1} show how large ($l=30$ mm) lengthscales overestimate the level of compressive stress between the two central joints. Intermediate lengthscales ($7 < l < 15$ mm) exacerbate the issue. As the lengthscale is further reduced ($l < 1.5$ mm) compressive stresses begin to diminish, and the stress profile develops two distinct (more compressive) troughs that flank a less compressive peak at the origin. Reducing the lengthscale causes similar changes to the vertical stress between two joints, which develops a distinct (tensile) peak in stress at the origin flanked by steep compressive stresses close to the fractures. Although not shown, the same phenomenon was observed for the AT2 model. We consider it likely that reducing the lengthscale further would allow the solution to converge towards the FDEM solution predicted by Bai and Pollard~\cite{bai_fracture_2000}.Unfortunately, such simulations became prohibitively expensive to run for very small length scales such that we could not explore the convergence behaviour. Therefore, at this stage it has to remain open at what lengthscale a “natural” stress distribution may emerge.

The temptative conclusions of the results above illustrate the challenges present in phase-field models for brittle fracturing. In essence, the regularisation of a fracture with the phase-field lengthscale significantly distorts the surrounding stress field both qualititively and quantitatively. Essentially, larger lengthscales result in stress profiles of joints that seem closer together than their actual spacing. The distortion is significant even when the lengthscale only occupies less than $\lesssim 1\%$ of the spacing between fractures. Such a result may explain the overestimation of fracture spacings reported in \cref{sec:pf_perspective}.

\subsection{Premature failure in compression and shear}
\begin{figure}[h]
\centering
  \includegraphics[width=0.85\linewidth]{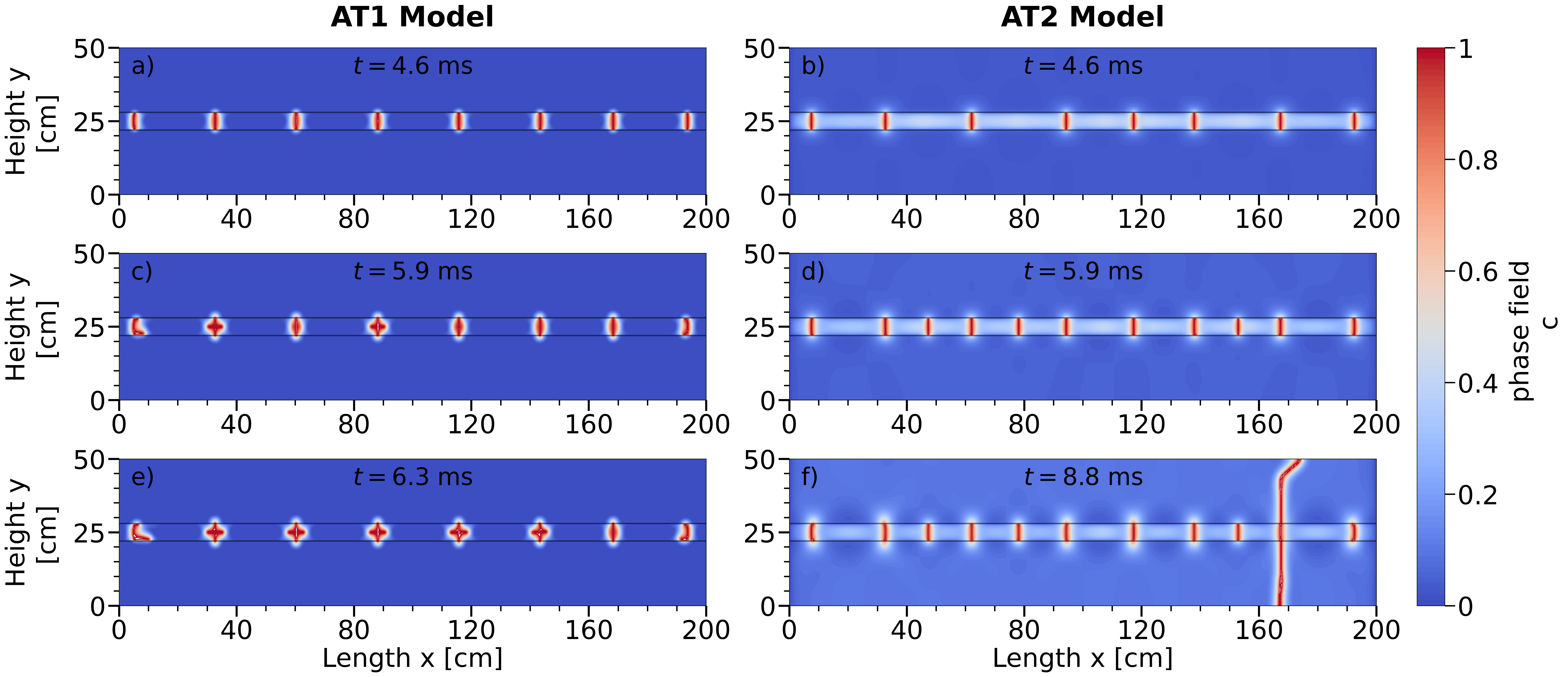}
  \caption{Premature initation of a compressive fracture in the AT1 model. Results showing the phase-field $c$ for the AT1 (left) and AT2 (right) model with parameters used from \cref{tab:mech_params}, with the exception of the shale fracture toughness set to $G_c=250$ J/m. Various snapshots in time $t$ are shown, with applied displacement $U=1$ cm. }
  \label{fig:AT_PhaseComp}
\end{figure}

In this section, we analyse how standard phase-field models induce rock in sedimentary layers to fail prematurely under compression and shear. Compressive failure is not observed within jointed sedimentary sequences, while shearing and delamination are common  \cite{helgeson_characteristics_1991}. In this context, `premature' failure  refers to failure that occurs at fracture spacings much greater than the height of the middle dolostone layer $(S/H_d > 1)$. Nonphysical fracturing under compression is observed when utilising the volumetric-deviatoric split alongside the AT1 model, while premature shearing and delamination at layer interfaces occurs instead with the AT2 model. To reproduce such results, the stress intensity factor of the surrounding shale layers are set $G_c^s\geq250$ J/m compared to the reference case. An important remark is that the nonphysical fracturing observed in this section occurs once `joint saturation' (no new infilling of joints) has been reached, and therefore, results herein are not independent to the unnatural stress state shown to be induced by the phase-field lengthscale in \cref{sec:stress_distorting_lengthscale}. 

The results in \cref{fig:AT_PhaseComp} are obtained using a shale fracture toughness $G_c^s=250$ J/m. The AT1 model predicts the propagation of an orthogonal fracture set that initiates from the joints center where the stress is compressive (seen in \cref{fig:AT_PhaseComp}c,e). No new joints are formed once compressive fractures propagate in the AT1 model, fixing the fracture spacing to approximately four times the dolostone layer height $S\sim4H_d$. The AT2 model also experiences a `saturation' of joints at a similar time ($t\sim5.9\,$ms) to the AT1 model (\cref{fig:AT_PhaseComp}d,f). However, in comparison to the AT1 model, the AT2 model accommodates the additional strain by propagating into the shale layer, eventually causing total failure of the specimen; a feature likely due to the difficulty the AT2 model has in capturing sharp interfaces \cite{nguyen_role_2019,yoshioka_variational_2021}. 
\hfill\newline

\begin{figure}[h]
\centering
  \includegraphics[width=1\linewidth]{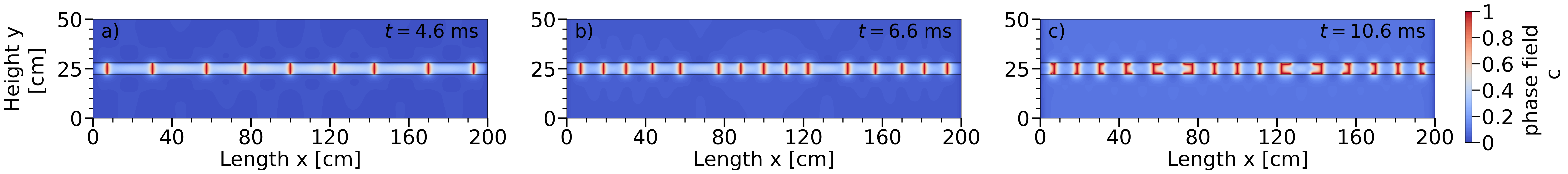}
  \caption{Phase-field $c$ distributions for AT2 model at various snapshots in time, with the fracture toughness in shale set to $G_c^s=400$ J/m.  }
  \label{fig:pf_AT2_gc400}
\end{figure}

Further increasing the shale fracture toughness ($G_c^s=400\,$ J/m) leads to the shearing and delamination of the layer interface for the AT2 model. Shearing can be seen to clearly occur in \cref{fig:pf_AT2_gc400} at $t=10.6$ ms. Despite the shale being rendered essentially impenetrable, ratios of fracture spacing to layer height remain greater than $2$ for the AT2 model. Such a result further corroborates the findings in \cref{sec:stress_distorting_lengthscale} that the phase-field lengthscale is inhibiting further infilling of fractures. Such shearing behaviour is also reproduced by the simulation results of Chukwudozie \textit{et al} \cite{chukwudozie_new_2013}.
\hfill\newline

An analysis of nucleation under multi-axial stress states \cite{de_lorenzis_nucleation_2022} offers an explanation for the compressive fracturing observed herein.  In phase-field models, the tensile, compressive, and shear strength of the material depends on the degradation and dissipation function, the energy-split, and any additional driving forces introduced in the formulation \cite{kumar_revisiting_2020,de_lorenzis_nucleation_2022}. For rocks, the compressive strength is pressure-sensitive, and up to a magnitude greater than it's tensile strength; the Drucker-Prager strength envelope being a classical example applied for rocks \cite{drucker_soil_1952}. For the simulations herein, the volumetric-deviatoric split employed in \cref{eq:vol_dev_split} results in a compressive ($\sigma_c$) and shear ($\sigma_s$) strength 
\begin{equation}
    \sigma_c=\sqrt{\frac{3}{2(1+\nu)}}\sigma_t\, , \qquad 
    \sigma_s = \frac{1}{\sqrt{3}} \sigma_c \, , 
\end{equation}
where the tensile strength $\sigma_t$ depends on which AT formulation is used (\cref{eq:tensile_strength}) \cite{de_lorenzis_nucleation_2022}. Consequently, using the parameters in \cref{tab:mech_params}, the compressive strength is predicted to be  $\sigma_c^{AT1} \sim 5.5\,$MPa for the AT1 model and $\sigma_c^{AT2}\sim3\,$MPa for the AT2 model. Furthermore, the shear strength is close to half the compressive strength ($\sigma_s\sim0.57\sigma_c$) according to the phase-field model. Such values are extremely small for rocks \cite{bahrami_theory_2020}. The spectral energy split could alleviate such issues since compressive and shear strengths are higher than the volumetric-deviatoric split ($\sigma_c\sim3\sigma_t$ for $\nu=0.25$) \cite{de_lorenzis_nucleation_2022}. However, such models deliver undesired residual crack-like stresses under opening or shearing \cite{vicentini_energy_2023}, and are therefore not investigated herein.
\hfill\newline

\begin{figure}[h]
\centering
  \includegraphics[width=0.7\linewidth]{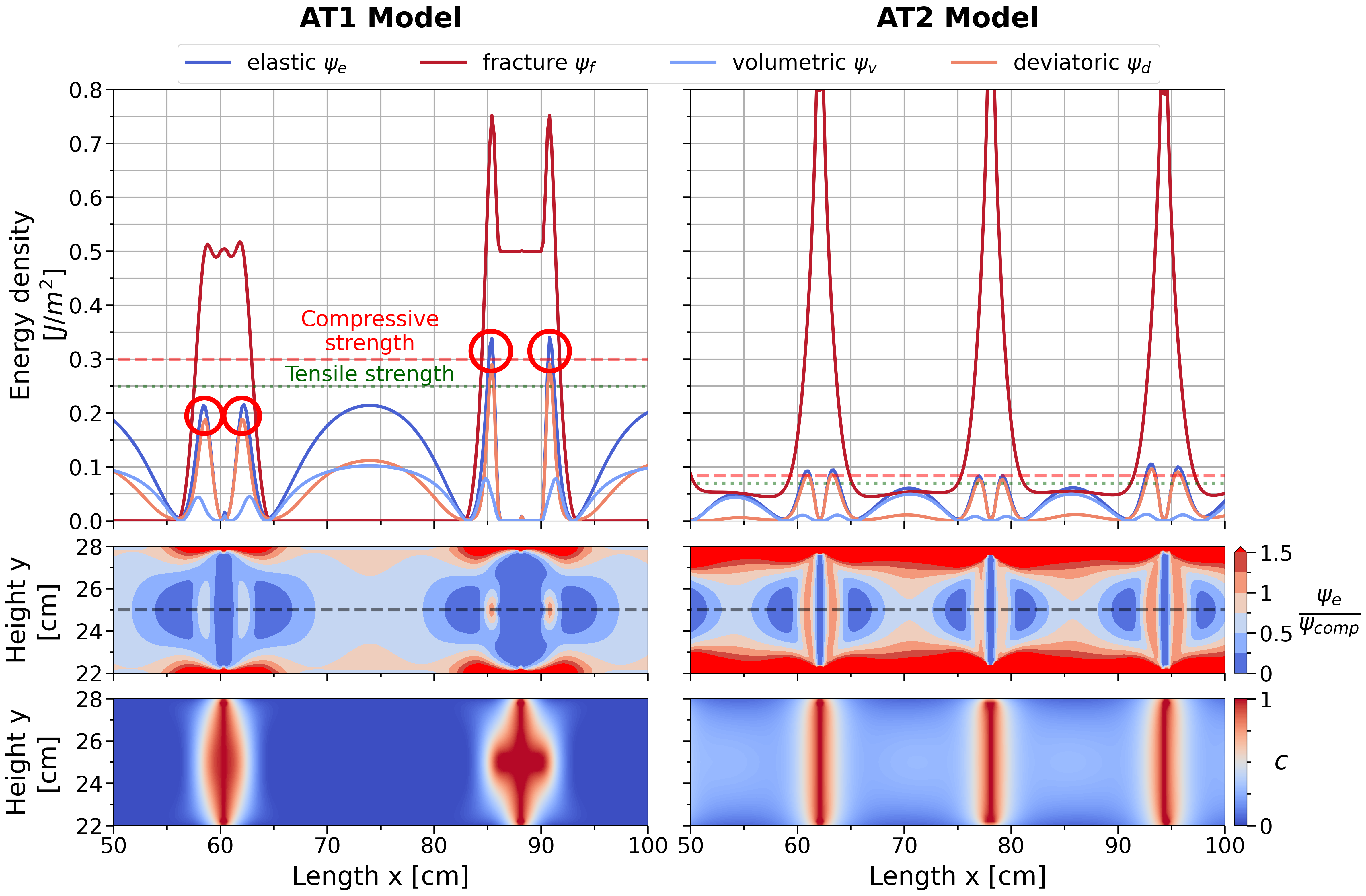}
  \caption{Results showing the energy and phase-field distribution for the middle dolostone layer. Both the AT1 (left) and AT2 (right) models are compared, with the time for each simulation taken significantly after joint saturation has been reached. The fracture toughness of the shale is $G_c^s=300$ J/m.  }
  \label{fig:CompEnergyDensity_ATCOMP}
\end{figure}

A closer look at the energy distribution within the dolostone layer confirms the diagnosis of premature compressive fracturing. The results on the left of  \cref{fig:CompEnergyDensity_ATCOMP} highlight the time-step after the AT1 model predicts a compressive fracture to brutally initiate from the middle of an existing joint. The compressive fracture can be identified by the peaks in deviatoric energy density, which cause the elastic energy density to surpass the compressive strength of the rock. Surprisingly, the deviatoric energy densities are also large for the AT2 model, surpassing the compressive strength of the rock in a refined neighbourhood around the fracture. However, for the AT2 model, an extensive area of rock experiences elevated shear stresses at the interface between the two sedimentary layers, therefore promoting the occurrence of shear and delamination before the onset of compressive failure like in the AT1 model. Since the AT2 model favours the penetration of the shale interface (as shown in \cref{fig:AT_PhaseComp}), inhibiting such penetration with a large shale fracture toughness naturally leads to the development of larger stresses at the interface. 
\hfill\newline

In general, both phase-field formulations lead to non-physical fracturing as the fracture toughness of the surrounding shale layers is increased. Such failure occurs after joint saturation has been reached, and is therefore dependent on the stress-distorting effects of the phase-field lengthscale explored in \cref{sec:stress_distorting_lengthscale}. By employing the volumetric-deviatoric split, compressive and shear strengths are set unrealistically low for rocks, causing unnatural fracture patterns to manifest as a result. The AT1 model favours the formation of unrealistic compressive fractures, while the AT2 model favours premature shearing at layer interfaces. In all cases, average fracture spacings are overestimated by at least a factor of 2 compared to outcrop observations. Overall, the volumetric-deviatoric phase-field formulation is essentially unsuitable for a quantitative analysis of fracturing in geological problems if both large tensile and compressive stresses develop in the rock; an unavoidable feature shown to be the case for layered sedimentary sequences subject to pure extension.     \section{Conclusion and Outlook}
\label{sec:conclusion}

\begin{figure}
\centering
\begin{subfigure}{0.45\textwidth}
    \includegraphics[width=\textwidth]{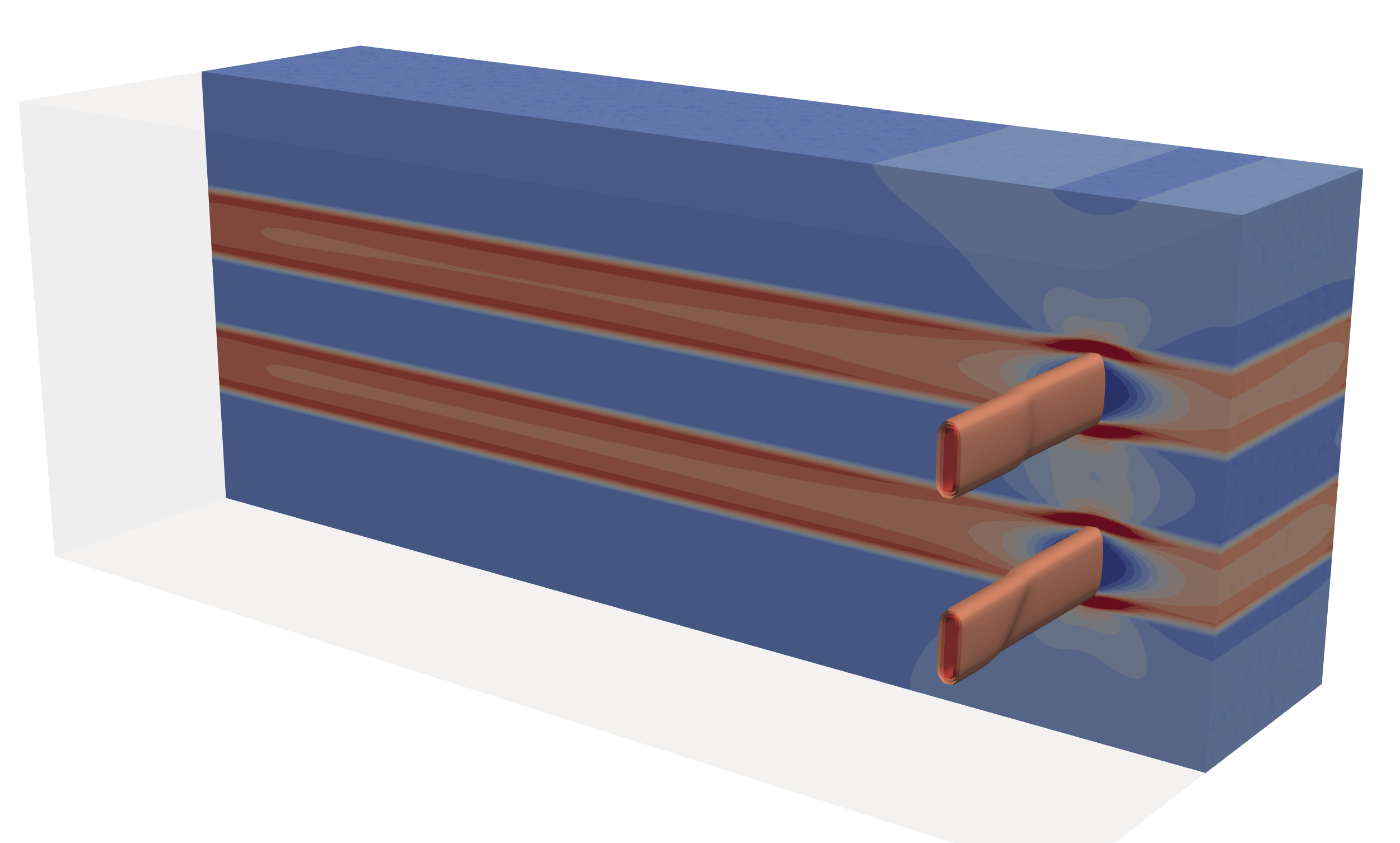}
\end{subfigure}
\hfill
\begin{subfigure}{0.45\textwidth}
    \includegraphics[width=\textwidth]{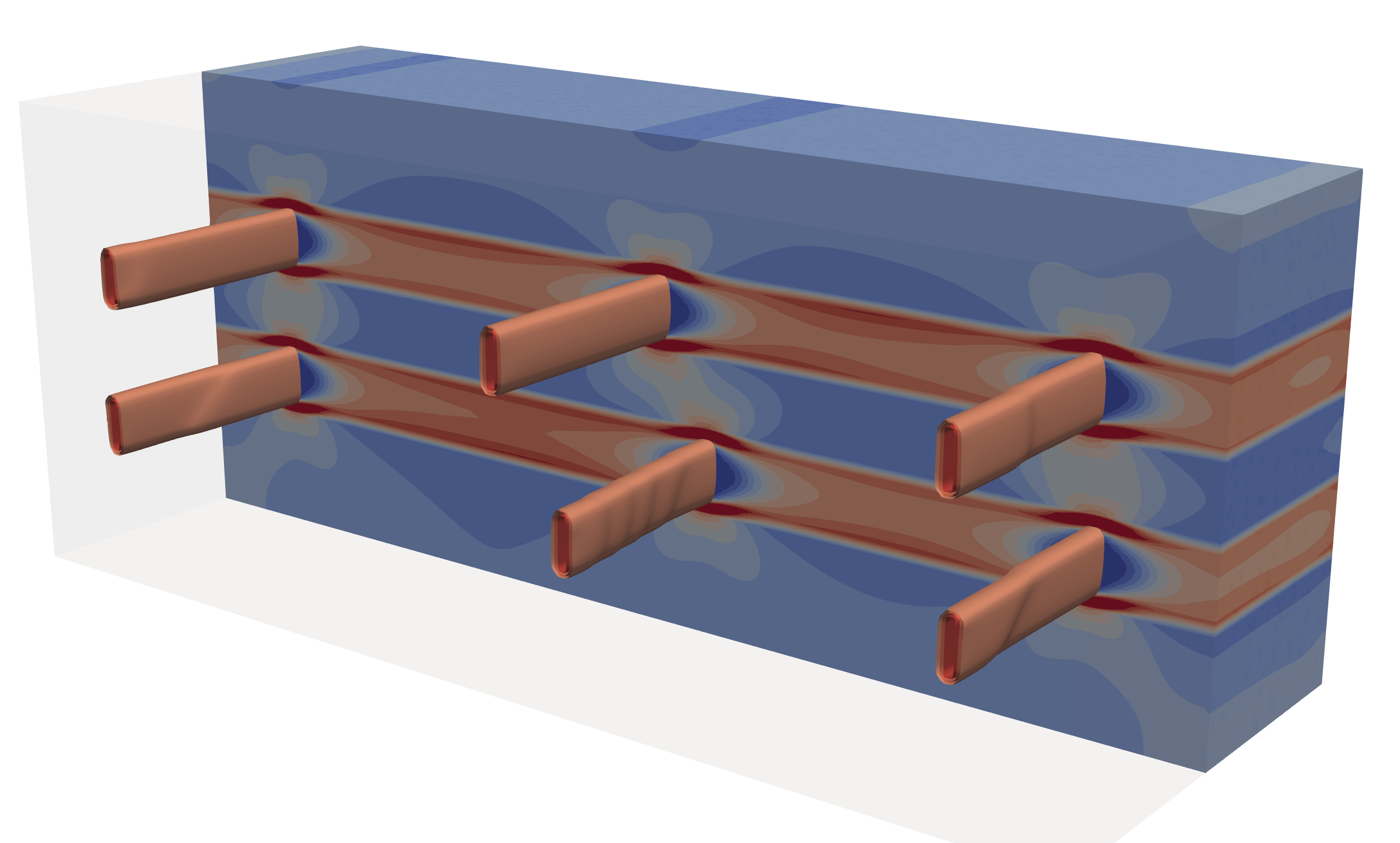}
\end{subfigure}
\hfill
\begin{subfigure}{0.7\textwidth}
    \includegraphics[width=\textwidth]{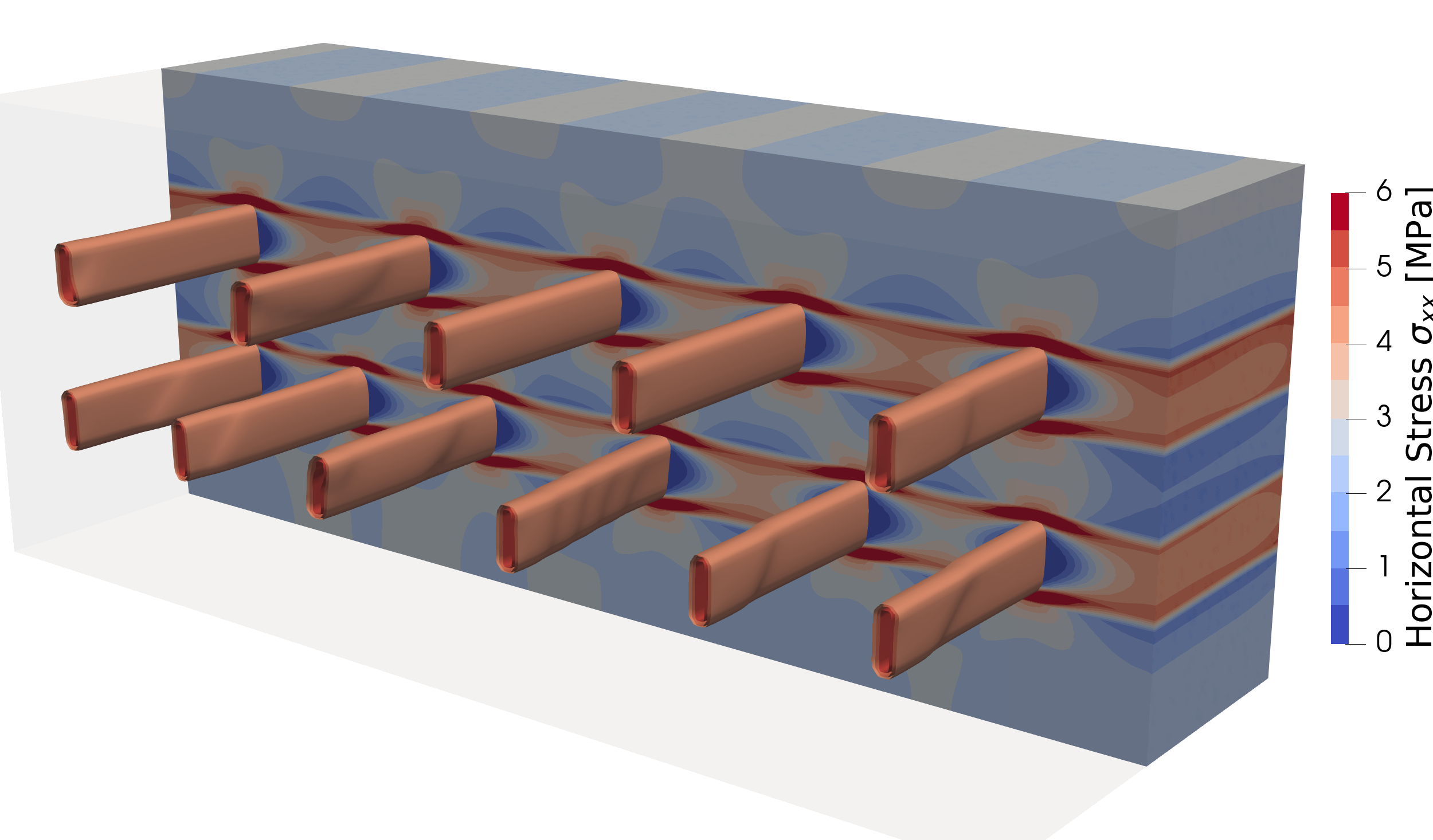}
\end{subfigure}
\caption{ 3 Dimensional phase-field and horizontal stress distribution within an extending five-layer sedimentary system using the AT2 model. Two stiff dolostone layers are interbedded between three compliant shale layers. Joints can be seen sequentially infilling along the isocontours for the phase-field at $c=0.95,\,0.85,\,0.75$, while their infilling is influenced by the stress communication between layers. As opposed to the material parameters in \cref{tab:mech_params}, the poisson ratio is kept at $\nu=0.25$ for all layers, while the shale fracture toughness is $G_c^s=300\,$ J/m.}
\label{fig:3DJointing}
\end{figure}

In this work, we studied the ability in reproducing jointing in sedimentary layers using the volumetric-deviatoric split alongside the AT1 and AT2 phase-field models. On the one hand, such standard phase-field models show promising features that qualitatively reproduce the jointing process. In particular, the models balance fracture nucleation and propagation elegantly in order to `organically' infill joints until joint saturation is reached. The models also qualitatively capture the relationship between joint spacing and the height of the stiffer sedimentary layer. Furthermore, phase-field models automatically balance the competing effects of jointing verses shearing within a single framework, providing a complete mechanical basis with which to study jointing; a feature lacking in alternative modelling approaches. Finally, stress interactions among multiple layers and in 3 dimensions are amenable to study, as is shown in \cref{fig:3DJointing}, where stress interactions cause joints to initiate at an offset to each other.   

On the other-hand, standard phase-field formulations are clearly shown herein to have significant limitations when predicting jointing. Standard phase-field models overestimate the spacing of joints by a factor of 2-4 when compared to most geological observations and predictions from damage mechanics \cite{bai_fracture_2000,li_fracture_2014}. The cause for the discrepancy was shown to be inherent to the stress-distorting effects of the phase-field lengthscale. Decreasing the lengthscale results in gradual improvement but becomes prohibitively computationally expensive for very small lengthscales such that the  stress profiles predicted by finite discrete element models (\cite{bai_fracture_2000}) was not reached in this study. Lengthscale effects are also suspected to cause challenges at sharp material interfaces, as shearing and delamination of layer interfaces is clearly favoured by the AT2 compared to the AT1 model.

The use of the volumetric-deviatoric energy split is shown to induce further limitations in predicting failure patterns when the fracture toughness of the shale is high. The cause of the problem lies in the unrealistically low compressive and shear strengths that are unavoidable when employing the volumetric-deviatoric energy split. Although a purely extensive stress is imposed on the specimen, unrealistic fracturing under compression occurs with the AT1 model, while premature shearing occurs with the AT2 model. Fortunately, the problem is likely surmountable with more recent formulations that attempt to modify the strength envelope in mixed loading conditions \cite{kumar_revisiting_2020,de_lorenzis_nucleation_2022,vicentini_energy_2023}.  This work greatly motivates their use when studying jointing, as compressive and shear stresses are shown to be unavoidable. 

Overall, the phase-field method shows promise in generating natural fracture networks, but significant work is needed if constitutive models are to handle failure under compression and shearing, alongside distortions induced by the phase-field lengthscale, including at material interfaces. Significant progress is already underway in this direction \cite{vicentini_energy_2023,yoshioka_variational_2021,zhou_interface-width-insensitive_2022,hansen-dorr_phase-field_2020}. This study shows the importance of benchmarking phase-field  constitutive models against geological observations to advance the technological readiness of the methodology.

\section*{Acknowledgement}
We would like to gratefully acknowledge the support of Platform for Advanced Scientific Computing (PASC)  through the project FraNetG:~Fracture Network Growth. Furthermore, we would like to heartfully thank the insightful and motivated discussions with Prof Laura De Lorenzis and Prof Keita Yoshioka.

\bibliography{biblio_EDDI_new.bib,extras}

\appendix
\section{Finite element discretisation}
\label{sec:FE_discretisation}

The use of the trust region method requires the discretised formulations of the energy function $f(\xv)$, its gradient $\nabla f$, and it's Hessian $\nabla^2 f$. 
The Galerkin finite element (FE) method is chosen herein for the discretisation, similar to that implemented in \cite{kopanicakova_recursive_2020,zulian_large_2021}, which uses a shape regular, quasi-uniform, conforming quadrilateral mesh $\pazocal{Q}_h$ of the domain $\Omega$. The nodal degrees of freedom $\hat{\uv}\in \R^{dN}$ and $\hat{\boldsymbol{c}} \in \R^N$ refer to the nodal displacement and phase-field variables, respectively. Consequently, the following finite element discretisation is used
\begin{equation}
    \label{eq:FE_approx}
   \uv(\xv) \approx \textbf{N}_u (\xv) \hat{\uv},  \hspace{1.0cm} \text{and} \hspace{1.0cm}  c(\xv) = \textbf{N}(\xv) \hat{\boldsymbol{c}},
  \end{equation}
  with bilinear nodal basis functions $\{N_i\}_{i=1}^n$, for $d=2$, such that 
\begin{equation}\label{eq:Shapefunctions}
   \textbf{N}_u =  \begin{bmatrix} N_1 & 0   & N_2 &  0  &  ... & N_n & 0 \\ 
                        0   & N_1 &  0  & N_2 &  ... & 0  & N_n 
    \end{bmatrix} \,, \qquad
    \textbf{N} =  \begin{bmatrix} N_1 & N_2 &  ... & N_n \\
                    \end{bmatrix}.
\end{equation}
As a result, the discretisation global energy $f(\hat{\uv},\hat{\boldsymbol{c}})$ can be written as   
\begin{linenomath*}
  \begin{equation}
    f(\hat{\uv}, \hat{\boldsymbol{c}}) = \int_{\Omega} g(\hat{\boldsymbol{c}})\psi^{d}_e(\hat{\uv}) + \psi^{r}_e(\hat{\uv}) \, d\Omega + G_c \int_\Omega \gamma(\hat{\boldsymbol{c}},\nabla N \hat{\boldsymbol{c}}) \,  d\Omega \, - \int_{\Gamma_N} \textbf{t} \cdot \uv d\Gamma \, .
    \label{eq:discretised_energy_regularized}
  \end{equation}
\end{linenomath*}
Discretising the strong form of the governing equations \cref{eq:strongform} gives an expression for the gradient $\nabla f = F(\hat{\uv}, \hat{\boldsymbol{c}})$ which takes the form
\begin{equation}
    \boldsymbol{F}(\hat{\uv},\hat{\boldsymbol{c}}) :=   
        \begin{bmatrix} \boldsymbol{F}_u(\hat{\uv},\hat{\boldsymbol{c}}) \\ \boldsymbol{F}_c(\hat{\uv},\hat{\boldsymbol{c}}) \end{bmatrix} = 
        \begin{bmatrix} \boldsymbol{0} \\ \boldsymbol{0} \end{bmatrix}\, , 
\end{equation}
with the following finite element residuals 
\begin{subequations}\label{eq:residuals}
\begin{equation}
    \boldsymbol{F}_u(\hat{\uv},\hat{\boldsymbol{c}}) = \int_\Omega \textbf{B}^T \textbf{D}(\hat{\boldsymbol{c}}) \textbf{B} \, d\Omega \, \hat{\uv} - \int_\Omega \textbf{N}_u^T b \, ,  
\end{equation}
\begin{equation}
    \boldsymbol{F}_c(\hat{\uv},\hat{\boldsymbol{c}}) = 2\int_\Omega \nabla\textbf{N}^T w_1l^2 \nabla \textbf{N} \, d\Omega \, \hat{\boldsymbol{c}} + \int \textbf{N}^T \,w_1 \omega'(\hat{\boldsymbol{c}}) \,d\Omega \, + \int_\Omega \textbf{N}^T \psi^d(\hat{\uv}) g'(\hat{\boldsymbol{c}})  \, d\Omega  \, , 
\end{equation}
\end{subequations}
where engineering (Voigt) notation is used throughout. Indeed, the elasticity matrix \textbf{D}$(\hat{\boldsymbol{c}})$ is used to represent the degraded elasticity tensor $C$ from \cref{eq:elasticity_tensor_C} in Voigt notation. The shape-function derivative matrices 
\begin{equation}\label{eq:BMatrices}
   \textbf{B} =  \begin{bmatrix}  N_{1,x} & 0       &  N_{2,x} &  0      &  ... & N_{n,x} & 0       \\ 
                                    0   & N_{1,y}     &  0       & N_{2,y} &  ... & 0       & N_{n,y} \\
                                    N_{1,y} & N_{1,x} &  N_{2,y} & N_{2,x} &  ... & N_{n,y} & N_{n,x} 
    \end{bmatrix} \, , \qquad
    \nabla\textbf{N} =  \begin{bmatrix} N_{1,x} & N_{2,x} &  ... & N_{n,x} \\
                                    N_{1,y} & N_{2,y} &  ... & N_{n,y}
    \end{bmatrix}
\end{equation}

The discretised Hessian $\boldsymbol{H} = \nabla^2 f$ is obtained by further differentiating the gradients $\boldsymbol{F}$ to give a block structure matrix $\boldsymbol{H} = \begin{bmatrix} \boldsymbol{H}_{uu} & \boldsymbol{H}_{uc} \\ \boldsymbol{H}_{cu} & \boldsymbol{H}_{cc}\end{bmatrix}$, where its components are given as
\begin{equation}
    \begin{aligned}
      \boldsymbol{H}_{uu} & =  \int_\Omega \textbf{B}^T \textbf{D}(\hat{\boldsymbol{c}}) \textbf{B} \, d\Omega \,,      \\
      \boldsymbol{H}_{uc} & = \int_\Omega \textbf{B}^T \textbf{D}'(\hat{\boldsymbol{c}}) \textbf{N} \, d\Omega\,,         \\
      \boldsymbol{H}_{cu} & = \int_\Omega \textbf{N}^T \textbf{D}'(\hat{\boldsymbol{c}}) \textbf{B} \, d\Omega\,,         \\
      \boldsymbol{H}_{cc} & =  2\int_\Omega \nabla\textbf{N}^T w_1l_s \nabla \textbf{N} \, d\Omega \,  + \int_\Omega \textbf{N}^T \psi^d(\hat{\uv}) g''(\hat{\boldsymbol{c}})\, d\Omega \, , 
    \end{aligned}
\end{equation}
where $\textbf{D}'(\hat{\boldsymbol{c}})$ is the derivative with respect to $\hat{\boldsymbol{c}}$ of the degraded elasticity matrix $\textbf{D}$. Similar derivations of the Hessian can be found in \cite{singh_fracture-controlled_2016}, although an alternative stress-split is employed in their case. 

In practice, the off-diagonal block $\boldsymbol{H}_{uc}$ is often omitted in the construction of the Hessian, as it has been found to hinder the convergence of nonlinear solvers~\cite{wick2017modified}. 
However, it is important to note that removing only the block $\boldsymbol{H}_{uc}$ results in a non-symmetric Hessian, which poses challenges for the convergence of linear iterative methods. 
Consequently, in our numerical simulations, we eliminate both coupling terms, $\boldsymbol{H}_{uc}$ and $\boldsymbol{H}_{cu}$, which allows us to employ the Conjugate Gradient (CG) method, preconditioned with algebraic multigrid, to solve the arising linear systems.

The penalty method was used to enforce irreversibility and boundedness of the phase-field. Enforcing boundedness is needed to eliminate negative phase-field values appearing when using the AT1 formulation described in \cref{eq:AT1}. In practice, only the lower bound constraint is penalised, where the upper bound on $c$ is enforced as a dirichlet condition once the crack reaches $c=1$. A finite element penalisation is implementation by adding the following terms to the energy function $f$:
\begin{subequations}

       \begin{equation}
        f \leftarrow f + \int_\Omega \frac{\gamma_{irr}}{2} \big(c^n - c^{n-1}\big)^2  H^-(c^n - c^{n-1}) \, d\Omega\,  + \,  \int_\Omega \frac{\gamma_{bou}}{2}  \big( c^n \big)^2 H^-(c^n) \, d\Omega\,,   
    \end{equation}
where $c$ is evaluated at the integration points using the finite element discretisation in \cref{eq:FE_approx}. The heaviside function $H^-(\cdot)$ is used to apply the penalty only where the constraint is violated. Similarly the gradient term $\boldsymbol{F}_c$ and hessian $\boldsymbol{H}_{cc}$ matrices are modified accordingly
\begin{equation}
        \boldsymbol{F}_c \leftarrow \boldsymbol{F}_c +  \int_\Omega \gamma_{irr} \textbf{N}^T (c^n - c^{n-1})  H^-(c^n - c^{n-1})\,d\Omega\, + \,  \int_\Omega \gamma_{bou} \textbf{N}^T c^n H^-(c^n)\, d\Omega\,, 
    \end{equation}
\begin{equation}
        \boldsymbol{H}_{cc} \leftarrow \boldsymbol{H}_{cc} +  \int_\Omega \gamma_{irr} \textbf{N}^T \textbf{N} H^-(c^n - c^{n-1})\,d\Omega\, + \,  \int_\Omega \gamma_{bou} \textbf{N}^T \textbf{N} H^-(c^n) \, d\Omega\,,
    \end{equation}  
\end{subequations}
Separate penalty parameters $\gamma$ are used to enforce irreversibility ($\gamma_{irr}$) and boundedness ($\gamma_{bou}$), which are chosen according to Gerasimov and De Lorenzis \cite{gerasimov_penalization_2019}
\begin{subequations}
    \begin{equation}
        \gamma_{irr}  =    \frac{27 G_c}{64 l \varsigma_{irr}^2}  \,,
    \end{equation}
    \begin{equation}
        \gamma_{bou}  =    \frac{9 G_c}{64 l \varsigma^2_{bou}} \bigg(\frac{\Bar{L}}{l} - 2\bigg) \,,
    \end{equation}
\end{subequations}
where $\Bar{L}$ is taken as the average dimensions of the domain $\Omega$. 
In the simulations, the tolerances $\varsigma_{irr}$ and  $\varsigma_{bou}$ are chosen to lie between $0.05-0.01$, similar to that recommended by Gerasimov and De Lorenzis \cite{gerasimov_penalization_2019}.
In particular, we choose $\varsigma_{irr}=\varsigma_{bou}=0.02$, as those values were found to provide the phase-field values $c>-0.01$ throughout the simulation.

\subsection{Trust region (TR) algorithm}
Upon discretisation, the following optimisation problem can be formulated,
\begin{equation}\label{eq:constrained_minimisation}
\begin{aligned}
\min_{\hat{\xv} \in \R^{n}} \ f(\hat{\xv}),
\end{aligned}
\end{equation}
where $f: \R^{n} \rightarrow \R$ denotes the discretised form of the non-convex energy functional~\eqref{eq:energy_regularized}, and $\hat{\xv} =[\hat{\boldsymbol{c}},\hat{\uv}]^T \in \R^{(d+1)N}$ is the solution vector, which contains the nodal coefficients of the displacement $\hat{\uv}$, and phase-field $\hat{\boldsymbol{c}}$.

At each $i$-th iteration, the TR method approximates $f$ by a quadratic model $m_i$, generated around the current iterate~$\hat{\xv}_i$. 
This model is then minimised in order to obtain the search-direction~$\sv_i$ as follows
\begin{equation}
\begin{aligned}
\min_{\sv_i \in \R^{n}}   & \ m_i (\sv_i) = f(\hat{\xv}_i) + \langle \nabla f(\hat{\xv}_i), \sv_i \rangle + \frac{1}{2} \langle \sv_i , \nabla^2 f(\hat{\xv}_i) \sv_i \rangle, \\
& \text{subject to} \  \| \sv_i \| \leq \Delta_i,
\label{eq:tr_subproblem}
\end{aligned}
\end{equation}
where~$\Delta_i > 0$ is called trust region radius, which provides a constraint on the size of the search direction~$\sv_i$. 
Once the correction~$\sv_i$ is obtained, its quality has to be assessed. 
This is achieved by evaluating the trust region ratio as
\begin{align}
\rho = \frac{f(\hat{\xv}_i)-  f(\hat{\xv}_i + \sv_i)}{m_i (\boldsymbol{0}) - m_i(\sv_i)} = \frac{\text{actual reduction}}{\text{predicted reduction}},
\end{align}
which compares the decrease in the objective function~$f$ with the decrease in model~$m_i$. 
If $\rho > \eta_1$, where $\eta_1 \in (0, 1)$, then the correction~$\sv_i$ provides a sufficient decrease in $f$ and it can be therefore accepted, i.e., $\hat{\xv}_{i+1} = \hat{\xv}_i + \sv_i$. 
Moreover, the model $m_i$ is believed to approximate $f$, and therefore, it can be trusted in larger regions. Hence, we enlarge the trust region radius~$\Delta_i$, which in turn loosens the step-size constraint in~\eqref{eq:tr_subproblem}. However, if $\rho \leq \eta_1$, then the correction~$\sv_i$ has to be disposed, i.e., $\hat{\xv}_{i+1} = \hat{\xv}_i$. In this case, the model is not considered to be an adequate representation of~$f$ in the current trust region. Hence, the trust region radius has to be shrunk.

\section{Verification of phase-field implementation }
\label{sec:app_verification}

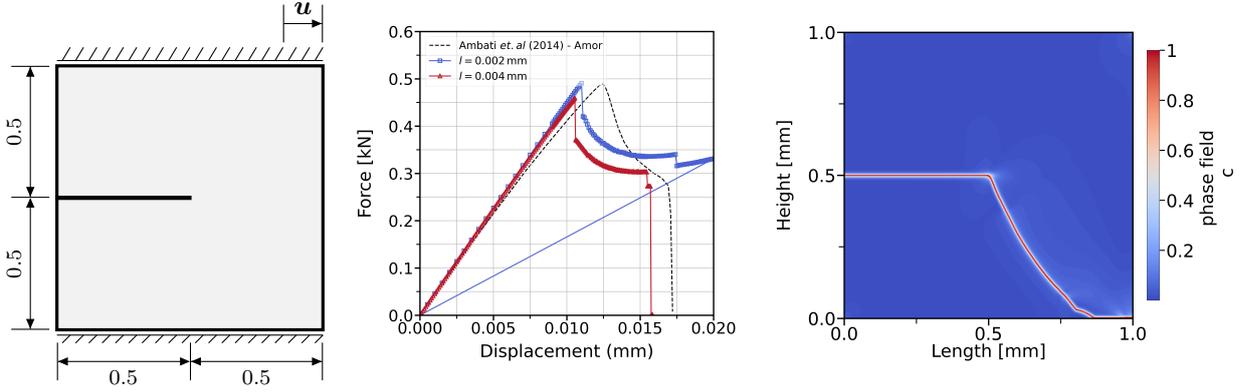
\begin{figure}[t]
\begin{centering}
  \hspace{5pt}
  \begin{minipage}{0.25\linewidth}
 \begin{tikzpicture}[scale=0.7]

      \draw[color=black, very thick, fill=gray!10!] (0.0, 0.0) rectangle (5.0, 5.0);
      \draw[color=black, very thick, fill=black] (0.0, 2.49) rectangle (2.5, 2.51);
      \draw[line width=0.5mm, color=black, thin] (0.0, -0.1) -- (5.0, -0.1);

      \draw[thin] \foreach \x [evaluate=\x as \y using \x/4]   in {1,...,20} { (\y, 5.35) -- (\y -0.15, 5.1) };
      \draw[line width=0.5mm, color=black, thin] (0.0, 5.1) -- (5.0, 5.1);

      \draw[thin] \foreach \x [evaluate=\x as \y using \x/4]   in {1,...,20} { (\y, -0.1) -- (\y -0.15, -0.25) };
      \draw[tipA-tipB, fill=black, thin](0,-0.6) -- node[below]{\footnotesize 0.5}  (2.5,-0.6);
      \draw[tipA-tipA, fill=black, thin](2.5,-0.6) -- node[below]{\footnotesize 0.5}  (5,-0.6);
      \draw[tipA-tipB, fill=black, thin](-0.5,0.0) -- node[above, rotate=90]{\footnotesize 0.5}  (-0.5, 2.5);
      \draw[tipA-tipA, fill=black, thin](-0.5,2.5) -- node[above, rotate=90]{\footnotesize 0.5}  (-0.5, 5.0);

      \draw[tipC-tipA, fill=black, thin ](4.25, 5.8) -- node[above]{$\uv$}  (5.0, 5.8);

\end{tikzpicture}   \end{minipage}
  \hspace{10pt}
  \begin{minipage}{0.3\linewidth}
    \includegraphics[height=4.5cm]{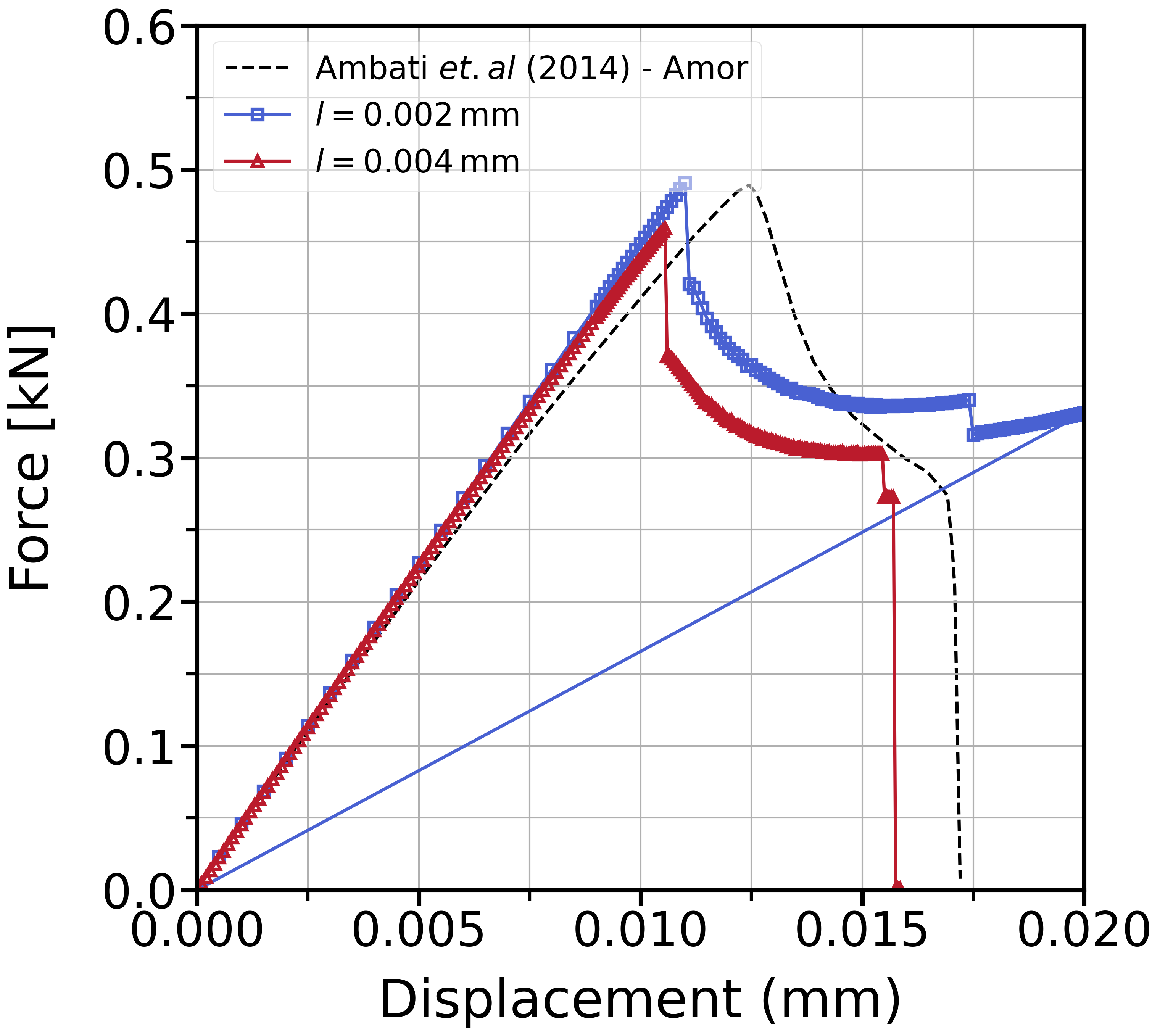}
  \end{minipage}
  \hspace{10pt}
  \begin{minipage}{0.3\linewidth}
    \includegraphics[height=4.5cm]{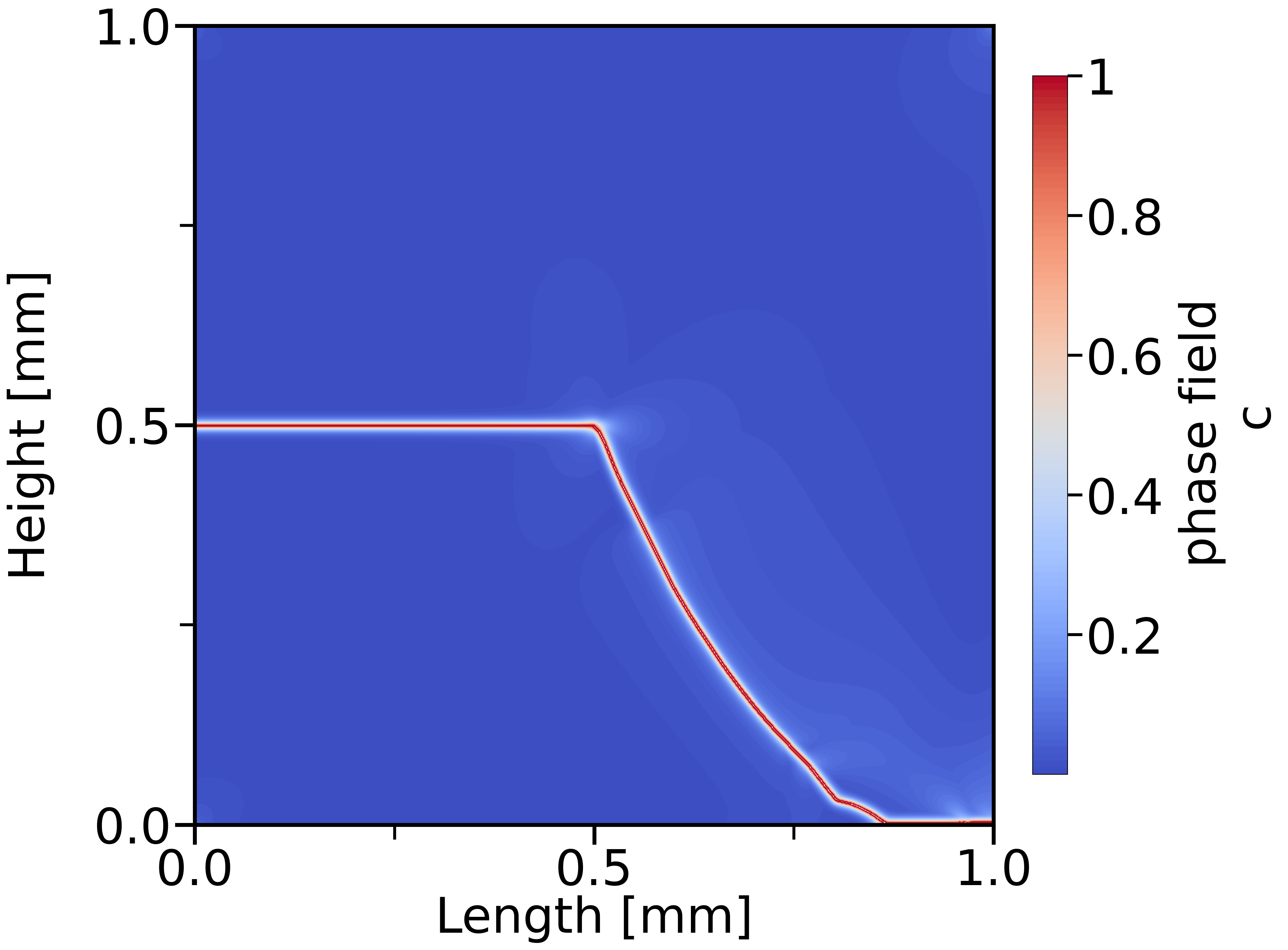}
  \end{minipage}
  \end{centering}
  \caption{Shear test: (Left) Geometry and boundary value problem setup, with units in $mm$. (Middle) Evolution of force displacement response of specimen, compared to Ambati et.al \cite{ambati_review_2015}. (Right) Final phase-field distribution upon rupture of the specimen }
  \label{fig:shear_setup_sim}
\end{figure}

The finite element implementation of the phase-field model herein has been validated in several occasions \cite{kopanicakova_recursive_2020, zulian_large_2021}. Nonetheless, herein we provide further verification for the problem of a single-edged notch under shear. Such a benchmark is commonly used to validate the implementation of the energy split, which in this case is the volumetric-deviatoric split. The bottom of the specimen is held fixed, while an incremental horizontal displacement is applied as shown in \Cref{fig:shear_setup_sim}. The dirichlet boundary conditions applied on the top boundary are $\uv_x := t \bar{u}$, where $\bar{u} = 1\,\text{mm/s}$, with pseudo-timesteps $\Delta t=10^{-4}$ for the first $0.009$ seconds, and then $\Delta t = 5 \times 10^{-5}\,s$ until complete failure occurs. A uniform resolution of $h=0.00125\,$mm was chosen, which is adequate for capturing the length-scale of the damage zone which is proportional to $0.004$. 

The results are shown in \cref{fig:shear_setup_sim} and compared to Ambati \textit{et.al} \cite{ambati_review_2015}. The material parameters are $\lambda=121.15\,$kN/mm$^2$, $\mu=80.77\,$kN/mm$^2$,  $G_c=2.7\times10^{-3}\,$kN/mm. Results are shown for $l=0.004$ and $l=0.002\,$mm, where the latter is equal to the lengthscale used in Ambati \textit{et. al}, due to the slightly different phase-field notation. Overall, the models reproduce the same peak strength. However, the results herein reproduce a sharp drop in stress upon failure, while results from Ambati show a smoothed out failure response. The difference is attributed to the strong non-linearity in the volumetric deviatoric split, which is adequately captured by the monolithic solver, which converges to very low residuals after reaching thousands of iterations. On the other-hand, Ambati uses a staggered solver which is known to exhibit a slow convergence, and utilises a fixed number of iterations.

The results of the shear test are verified against the numerical results of \cite{ambati_review_2015}. The material parameters are $\lambda=121.15\,$kN/mm$^2$, $\mu=80.77\,$kN/mm$^2$,  $G_c=2.7\times10^{-3}\,$kN/mm, and $l=0.002\,$mm, which is half the lengthscale used in Ambati \textit{et. al}~\cite{ambati_review_2015} due to the slightly different phase-field formulations.

\end{sloppypar}

\end{document}